\documentclass[twocolumn]{aastex63}
\received{February 20, 2025}
\revised{May 27, 2025}
\accepted{\today}
\submitjournal{AJ}
\shorttitle{Near-IR observations characterization of IRAS 16475--4609}
\shortauthors{Navarete et al.}
\usepackage{graphicx}	% Including figure files
\usepackage{enumitem}
\usepackage{todonotes}
\usepackage{xspace}
\newcommand{\hh}{H$_2$\xspace}
\newcommand{\brg}{Br$\gamma$\xspace}

\newcommand{\trot}{$T_{\rm{rot}}$\xspace}
\newcommand{\ntot}{$N_{\rm{tot}}$\xspace}
\newcommand{\ak}{$A_{\rm{Ks}}$\xspace}
\newcommand{\iras}{$\textrm{IRAS~16475$-$4609}$\xspace}
\newcommand{\uchii}{$\textrm{UC\ion{H}{2}}$\xspace}
\newcommand{\hii}{$\textrm{\ion{H}{2}}$\xspace}
\newcommand{\vlsr}{V$_{\textrm{lsr}}$\xspace}

\let\oldAA\AA
\renewcommand{\AA}{\oldAA\xspace}

\begin{document}

%\title{IRAS 16475--4609: a Young Compact HII Region Seen in the Near-Infrared}

\title{IRAS 16475--4609: A Young Compact HII Region Sculpting Its Molecular Environment}

\correspondingauthor{Felipe Navarete}
\email{fnavarete@lna.br}

\author[0000-0002-0284-0578]{Felipe Navarete}
\affiliation{Laboratório Nacional de Astrofísica, Rua dos Estados Unidos 154, 37504-364, Itajubá, MG, Brazil}
\affiliation{NSF's NOIRLab/SOAR Telescope, Casilla 603, La Serena, 1700000, Chile}

\author[0000-0002-4596-1337]{Sean D. Points}
\affiliation{NSF’s NOIRLab/CTIO, Casilla 603, La Serena, 1700000, Chile}

\author[0000-0002-7978-2994]{Augusto Damineli}
\affiliation{Instituto de Astronomia, Geof\'isica e Ci\^encias Atmosf\'ericas da USP, Rua do Mat\~ao 1226, Cidade Universit\'aria, S\~ao Paulo, Brasil}

%% Mark off the abstract in the ``abstract'' environment. 
\begin{abstract}
{We present a near-infrared spectroscopic and imaging analysis of the star-forming region \iras, based on TripleSpec/SOAR spectroscopy and NEWFIRM/CTIO imaging, complemented by archival radio and sub-millimeter data.
Our spectroscopic analysis indicates that the central source is an early B-type star (B0-B0.7~V) powering a compact \hii region characterized by strong \ion{H}{1} and \ion{He}{1} recombination lines, and molecular \hh emission. 
We derive a distance of 3.51$\pm$0.74~kpc, consistent with the position of the Scutum-Crux near arm at Galactic longitudes of $\sim$340$^\circ$.
At this distance, the ionized gas traced by \brg emission has a radius of 0.27$\pm$0.06~pc, placing the source in a transition phase between ultra-compact and compact \hii regions.
From radio data, we estimate an ionizing photon flux of {N$_{ly}$~=~(2.3$\pm$0.3)$\times10^{47}$~photons~s$^{-1}$}, and an electron temperature of {T$_e$\,=\,(5.4$\pm$0.2)$\times$10$^3$~K} for the ionized gas.
The analysis also reveals an obscured high-density molecular clump southwest of the \hii region, coincident with an ATLASGAL sub-millimeter peak, indicating a potential site of ongoing and triggered star formation as the ionization front advances into the surrounding molecular material.
These results suggest that \iras is a young high-mass star-forming region with stellar feedback actively shaping its environment, offering valuable insight into the early evolution of compact \hii regions.}
%
%%%%%%%%%%%%%%%%%%%%%%%
%
%We present the analysis of \iras through near-infrared TripleSpec/SOAR spectroscopic and complementary NEWFIRM/CTIO imaging observations.
%
%Our findings indicate the central source is an early B-type star (B0-B0.7~V) {powering} a {compact} \hii region characterized by strong \ion{H}{1} and \ion{He}{1} {recombination lines, along with molecular \hh emission}. 
%
%We derived a weighted mean distance of 3.51$\pm$0.74~kpc, consistent with the position of the Scutum-Crux near arm at Galactic longitudes of $\sim$340$^\circ$.
%At this distance, the extent of the nearly spherical ionized gas traced by the \brg emission has a radius of 0.27$\pm$0.06~pc, {placing the source in a transitioning phase between ultra-compact and compact \hii regions}. 
%
%{From radio continuum data, we estimate an ionizing photon flux of {N$_{ly}$~=~(2.3$\pm$0.3)$\times10^{47}$~photons~s$^{-1}$}, and derive an electron temperature of {T$_e$\,=\,(5.4$\pm$0.2)$\times$10$^3$~K} for the ionized gas.} 
%
%Our analysis also reveals an obscured high-density region towards the southwest direction of the \hii region, supported by ancillary sub-mm observations from ATLASGAL, suggesting potential sites for triggered star formation as the ionization front advances into the surrounding molecular material.

\end{abstract}

%% Keywords should appear after the \end{abstract} command. 
%% See the online documentation for the full list of available subject
%% keywords and the rules for their use.

\section{Introduction}
\label{sec_intro}

High-mass star formation {plays} a fundamental {role} in shaping the structure and evolution of the interstellar medium (ISM).
Throughout their lives, these stars significantly impact their surroundings, emitting intense ultraviolet (UV) radiation, driving powerful stellar winds, and eventually exploding as supernovae, providing feedback mechanisms for subsequent and triggered star formation \citep{Lada87} and the overall dynamics of the Galaxy.
During their early stages, high-mass stars ionize their immediate environment, creating compact \hii regions within the dense molecular clouds from which they formed \citep{Wood89}. 
\hii regions are not only crucial for understanding the physical conditions around young stars but are also among the most prominent tracers of star-forming regions and the spiral structure of the Galaxy \citep{Moises11,Hou14}.

{In particular, the ultra-compact \hii region (\uchii) represents the earliest observable phase of this process, characterized by small sizes ($\leq$~0.1~pc), high electron densities (10$^4$–10$^5$~cm$^{-3}$), and strong free-free radio emission \citep{Kurtz05}.
As the spectral energy distribution of UC\ion{H}{2}s peaks in the far-infrared (FIR), they have historically been identified through their characteristic FIR colors from large-scale surveys such as IRAS and Spitzer \citep{Wood89, Mottram10}.
Indeed, \citet{Wood89b} identified \uchii candidates based on FIR flux density (F$_\lambda$) ratio criteria, with log(F$_{25\mu m}$/F$_{12\mu m}$)~$>$~0.57 and log(F$_{60\mu m}$/F$_{12\mu m}$)~$>$~1.3, which serve as robust indicators of dense, dusty environments typically surrounding young high-mass stars.}

Despite the critical role of young \hii regions in high-mass star formation, many of these regions remain poorly characterized, limiting our understanding of their physical conditions and evolutionary states. One such example is \iras, a source that has been only superficially investigated in previous surveys.
Located at Galactic coordinates $\ell$=339\fdg72, $b$=--1\fdg20, slightly away from the densest regions of the Galactic plane and in the vicinity of well-known high-mass star-forming regions such as RCW~106 (d~=~3.5--4.0~kpc, \citealt{Russeil2003, Moises11}), G339.88--1.26 (d~=~2.8$\pm$0.5~kpc, \citealt{Green17}), and Westerlund~1 (d~=~4.05$\pm$0.20~kpc, \citealt{Navarete22}), \iras has the potential to provide insights into the interaction between molecular clouds and ionizing radiation from high-mass stars.

{The FIR colors of \iras -- log({F}$_{25\mu m}$/{F}$_{12\mu m}$)~=~0.7 and log({F}$_{60\mu m}$/{F}$_{12\mu m}$)~=~1.5 --} are consistent with the expected \uchii thresholds from \citet{Wood89b}, reinforcing its classification as a candidate \uchii region surrounded by a dense and dusty environment.
{The detection of 6.668-GHz methanol maser emission represents another key observational signature of the earliest phases of high-mass star formation, as such masers are typically associated with young, deeply embedded massive protostars \citep{Menten91,Urquhart13a}. For \iras, positive metanol maser detection was reported by \citet{Walsh95} but subsequent observations by \citet{Walsh97} did not confirm this emission, suggesting variability or a transient nature of the maser.}

Additionally, \iras has been identified as a diffuse \hii region (G339.7178$-$01.2089) in the Red MSX Source survey \citep{Lumsden13} and classified as a P-type object based on its low-resolution mid-infrared spectral properties \citep{Kwok97}. The presence of polycyclic aromatic hydrocarbon (PAH) emission features, characteristic of star-forming regions, points to a dense and dusty environment around the source, consistent with the conditions expected for a young \hii region.

While the combined evidence from FIR colors, methanol maser variability, and mid-infrared spectral properties suggests that \iras hosts a young, massive star embedded in a dense molecular cloud, these observations {lack the spatial and spectral resolution needed to resolve the detailed physical structure of the ionized and molecular gas. Thus, higher angular resolution is essential to disentangle compact ionized emission from surrounding extended nebular or continuum components, allowing a clearer characterization of the \hii region morphology and its interaction with the molecular environment. Furthermore, high spectral resolution enables accurate measurement of line profiles and radial velocities, which are crucial for determining a reliable distance estimate and, consequently, better constrain the location of \iras within the Galaxy.}

{To address these limitations, this} work presents a comprehensive analysis of \iras using near-infrared {(near-IR)} imaging and spectroscopic data to characterize its central star, the surrounding nebular emission, and its local ISM environment. By combining these observations {with archival data at longer wavelengths}, we aim to {constrain the} true nature of this source and {to} understand its role in the broader context of the {star formation process in} this region.

This manuscript is organized as follows. Section~\ref{sec_obs} outlines the observations and data processing.
The results from imaging and ancillary maps are presented in Sect.~\ref{sec_imaging}.
The near-IR spectroscopic analysis is presented in Sect.~\ref{sec_spectroscopy}, where we analyze the central source and the nebular gas. In Sect.~\ref{sec_distance}, we derive the distance to \iras and evaluate the physical parameters of the \hii region.
Finally, the main conclusions are discussed in Sect.~\ref{sec_discussion}.

\section{Observations and Data Reduction}
\label{sec_obs}

\subsection{NEWFIRM/CTIO}

Near-IR narrow and broad-band imaging observations were carried out using the wide-field near-IR camera NEWFIRM \citep{Autry03} installed on the 4-m Victor Blanco telescope at CTIO (Cerro Tololo, Chile) on July 18th, 2024. NEWFIRM has a detector mosaic of four 2k$\times$2k InSb arrays, providing a FOV of {27\farcm6$\times$27\farcm6} with a cross-shaped {35\arcsec}-wide gap between the arrays.

Observations were performed using the narrow-band \ion{Fe}{2} (1.644~{\micron}), \hh (2.144~{\micron}) and \brg (2.168~{\micron}), and the broad-band H and K filters. A four-point dithering pattern spaced by {82\arcsec} in a 2$\times$2 grid was adopted to suppress defects on the detector. The NEWFIRM observations are summarized in Table~\ref{table_newfirm}.

Basic data reduction (flat-field correction and sky subtraction) was performed using the IRAF MSCRED package. The astrometric solution and distortion correction of the field were obtained by comparing the field with reference stars in the 2MASS catalog using \texttt{SExtractor} and \texttt{Scamp} \citep{Bertin02}, leading to positional errors smaller than 0\farcs10. Finally, the individual images were median-combined using \texttt{Swarp} \citep{Bertin02}. 

The continuum subtraction of the narrow-band images was performed by scaling the broad-band image with a fixed factor, determined by minimizing the residuals of point sources across the FOV. This multiplicative factor was derived through an iterative process, adjusting the scale until the stellar residuals in the resulting narrow-band frame were minimized, ensuring that the scaled continuum contribution from the broad-band image was then subtracted from the narrow-band frame to isolate the emission-line features.

\setlength{\tabcolsep}{2pt}
\begin{table}[!ht]
    \centering
    \caption{Log of NEWFIRM/CTIO observations.}
    \label{table_newfirm}
    \begin{tabular}{c|ccccc}%cc}
    \hline
    \hline
Filter & $\lambda$/$\Delta\lambda$ & MJD & t$_{exp}$ & Airmass & FWHM           \\
       &  (nm)                      &    & (sec)     &         & (\arcsec)      \\
    \hline
\ion{Fe}{2}  & 1649.1/17.2 & 60479.235 & 16$\times$60 & 1.12 & 1.170$\pm$0.055 \\
\hh          & 2131.9/24.0 & 60479.242 & 16$\times$60 & 1.13 & 1.188$\pm$0.061 \\
\brg         & 2177.0/24.4 & 60479.253 & 17$\times$60 & 1.16 & 1.164$\pm$0.049 \\
H            & 1631/308    & 60479.273 & 12$\times$2  & 1.23 & 1.174$\pm$0.086 \\
K            & 2150/320    & 60479.267 & 13$\times$2  & 1.20 & 1.090$\pm$0.041 \\
    \hline
    \end{tabular}
\end{table}
\setlength{\tabcolsep}{6pt}

\subsection{Broad-band near-IR images from VVV}

We obtained ancillary near-IR images in the broad-band {$JHK_s$} filters taken with the VISTA telescope and part of the VISTA Variables in the Via Lactea (VVV) survey \citep{Minniti10}. The images have a plate scale of  {0.34\arcsec~pixel$^{-1}$} and median FWHM of {1.07$\pm$0.07\arcsec}.

\subsection{TripleSpec4.1/SOAR}

Spectroscopic observations were carried out with the TripleSpec~4.1 {near-IR} Imaging Spectrograph \citep[TripleSpec,][]{Schlawinl14} at the Southern Astrophysical Research Telescope (SOAR, Cerro Pach\'on, Chile). TripleSpec is a cross-dispersed spectrograph with a fixed {1.1$\times$28\arcsec} slit,  providing a full spectral coverage from 0.94 to 2.47~{\micron} (R$\sim$3,500), encompassing the entire $YJHK$ photometric range. 

Observations were carried in 2022 July 30th, with airmass of 1.92. We adopted an AABAA dither pattern with an exposure time of 150~sec per frame, where A and B correspond to a on- and off-source positions, respectively. The B position was set to a nearby nebular-free position. In order to map both the point source and the extended nebulosity, we adopted a position angle of 50$^\circ$, as indicated in Fig.~\ref{fig_images}. 

\subsection{Spectroscopic Data Reduction}
\label{sec_datared}

The TripleSpec/SOAR data was processed using a modified version of the IDL-based \texttt{Spextool} pipeline \citet{Cushing2004}, for use at SOAR. The data reduction steps include flat-field correction, wavelength calibration using a CuHeAr arc lamp, removal of emission sky features by subtracting A and B exposures, and extraction of the one-dimensional spectra. We used observations of the telluric standard star HIP~82091 observed at similar airmass values (1.84) to perform the telluric correction and flux calibration of the spectrum.
The final one-dimensional {near-IR} spectrum has a peak SNR of $\sim$300 at the center of the K-band. 

\subsubsection{Two-dimensional TripleSpec/SOAR spectroimages}
\label{sec_spectroimages_def}

Spectroimages of orders $n$~=~3 to 7 were obtained using the IDL \texttt{tspec\_spectro2d}\footnote{the IDL and Python versions of \texttt{tspec\_spectro2d} routine can be found at \url{https://github.com/navarete/}} procedure as described below.

Input raw data of on- ($ON$) and off-source ($OFF$) positions are divided by the flat-calibration frame created by \texttt{Spextool}. Then, the images are subtracted (producing a $ON_{\rm sub}$ frame), so the discrete telluric emission lines are suppressed from the on-source exposure. The user can improve the sky subtraction by modifying the multiplicative \texttt{off\_ratio} parameter (default value equals to one), defined as:
$$ON_{\rm sub} = ON - OFF \cdot {\rm off\_ratio}$$ 

The wavelength and spatial solutions obtained by \texttt{Spextool} were used to retificate the data in a linear grid for both spectral and spatial dimensions, by keeping the original plate scale (0.3376\arcsec~pixel$^{-1}$) and spectral dispersion of the instrument for each order.

Finally, the one-dimensional flux calibration solution obtained by the \texttt{xtelcorr} procedure was used to convert flux from DN~s$^{-1}$ to W~m$^{-2}$~\micron$^{-1}$ units.

\begin{figure*}[!ht]
    \centering
    {\includegraphics[width=\linewidth]{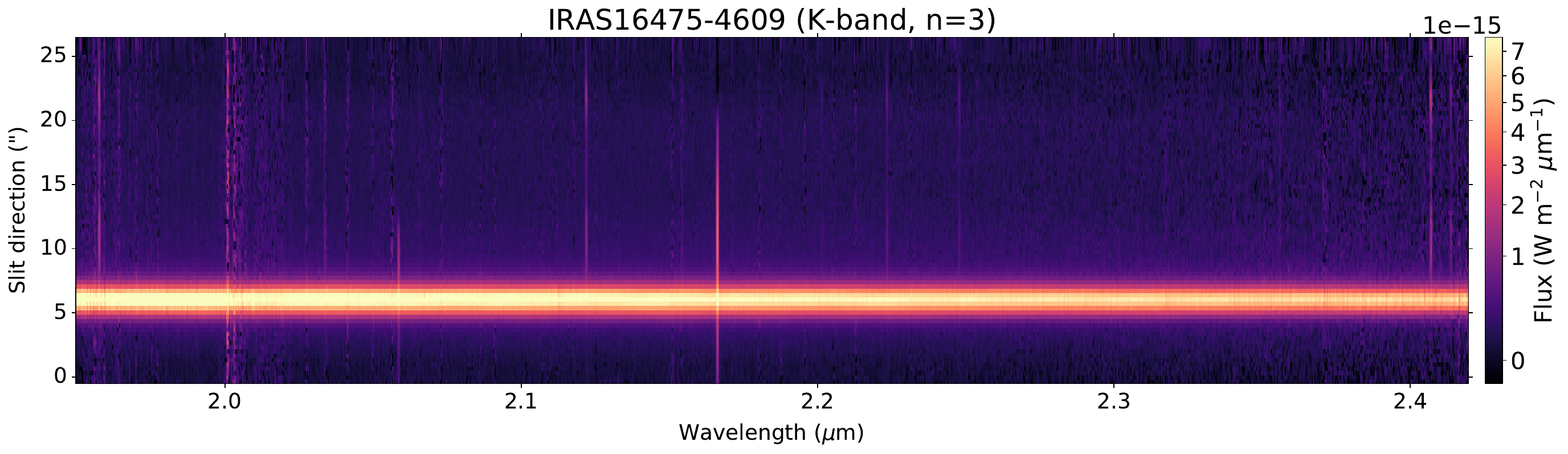}} \\[-1.0ex]
    \caption{Two-dimensional spectroimage of the order n=3 (K-band) resampled into a 2040$\times$80~pixel linear grid.}
    \label{fig_specimg_kband}
\end{figure*}

Final two-dimensional images of the four on-source exposures were median-combined, suppressing noise and cosmic rays. Figure~\ref{fig_specimg_kband} exhibits the spectroimage of order $n$~=~3 ($K$-band) as an example.

\section{The structure of IRAS 16475-4609}
\label{sec_imaging}

\begin{figure*}[!ht]
    \centering
    {\includegraphics[width=\linewidth]{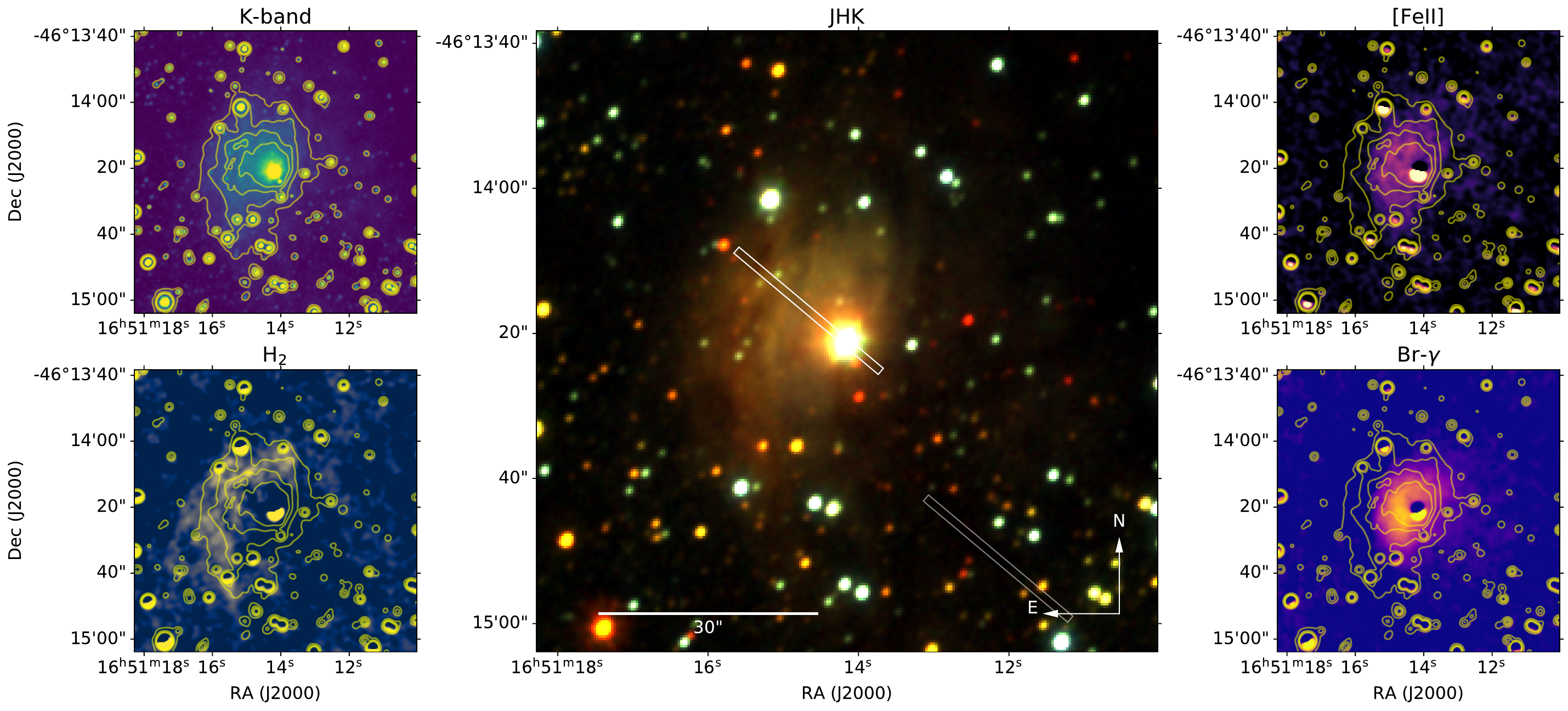}} \\[-1.0ex]
    \caption{NEWFIRM/CTIO and VVV/VISTA observations of \iras.
    The central panel shows a false-color RGB map {from VVV} (blue: J, green: H, red: {K$_s$}) of a {85$\times$85\arcsec} region centered at RA~=~16:51:14.156, Decl.~=~$-$46:14:21.23 (J2000). The white and grey rectangles indicate the A and B positions for the TripleSpec observations. A {30\arcsec} scale bar is indicated in the lower left region of the panel. 
    The top left panel shows the NEWFIRM K-band image of the same FOV, overlaid by contours highlighting the continuum emission at 15, 30, 45, and 60\% of the nebulosity peak intensity.
    The other panels show the continuum-subtracted \hh map (2.12~{\micron}, lower left), \ion{Fe}{2} (1.644~{\micron}, upper right), and \brg (2.166~{\micron}, lower right), {overlaid by the K-band contours}.
    }
    \label{fig_images}
\end{figure*}

In this section, we analyze the structural characteristics of \iras through near-infrared imaging, focusing on the interplay between the central source and its surrounding nebulosity.
Figure~\ref{fig_images} presents the {near-IR} maps obtained in both broad- and narrow-band filters. The false color $JHK_s$ map (central panel) reveals a nebulosity extending northeast (NE) from the bright source at the center of the FOV.
The nebular emission exhibits a conical shape, suggesting that the ionizing source has carved a cavity within a dense molecular cloud. Such interpretation is supported by the presence of deeply embedded near-infrared objects located to the south (S) and west (W) of the source, likely hidden within the dusty environment.

The $K$-band emission (top left panel) indicates that the continuum emission extends radially up to {$\sim$24\arcsec} from the central source, predominantly in the NE direction. In general, the $K$-band nebulosity is often associated with dust that has been heated by a nearby hot source.

The remaining panels in Fig.~\ref{fig_images} present continuum-subtracted narrow-band maps, exhibiting the emission of molecular hydrogen at {2.12~\micron} (\hh, lower left), the forbidden \ion{Fe}{2} line at 1.664~{\micron} (top right), and the \brg transition at 2.166~{\micron} (lower right).
The \brg and \ion{Fe}{2} maps probe the ionized gas in closer proximity to the central source. Notably, the unobstructed view to the NE direction displays a spherical morphology, characteristic of the ionization front typically generated by a massive star.
In contrast, the \hh emission traces the outer layer of the continuum nebulosity and the compact ionized gas, suggesting its location along the conical surface on the molecular cloud.

\begin{figure}[!ht]
    \centering
    {\includegraphics[width=\linewidth]{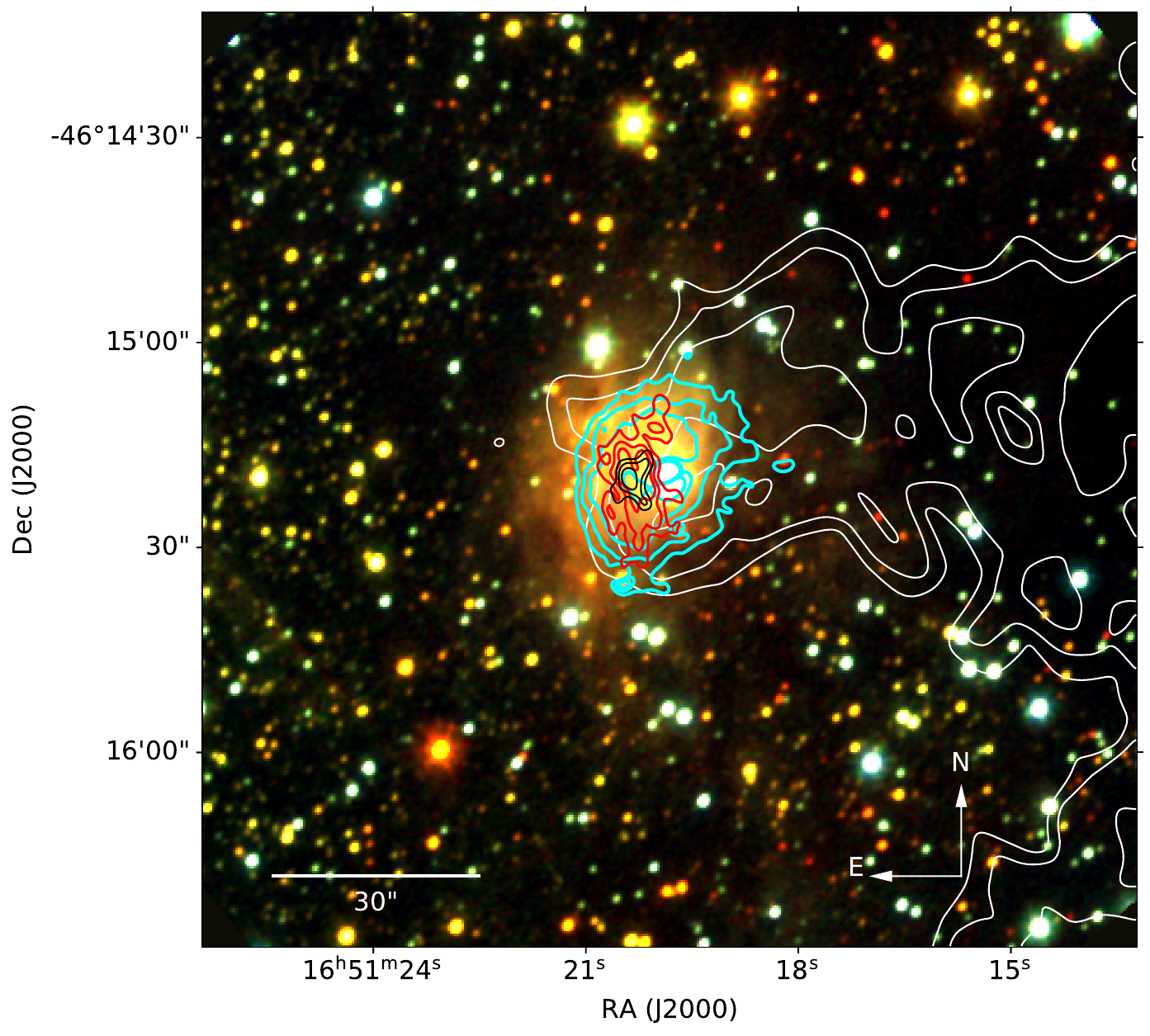}} \\[-1.0ex]
    \caption{VVV/VISTA observations of \iras in a {2\farcm25$\times$2\farcm25} region. The map shows a false-color RGB image highlighting the nebular emission around the IRAS object. A {30\arcsec} scale bar is indicated in the lower left region of the panel. The contours indicate the emission at 870~{\micron} from ATLASGAL (white curves), the ATCA maps at 3~cm (red) and 6~cm (black), and the near-IR \brg emission (cyan).}
    \label{fig_images_submm}
\end{figure}

The hypothesis of a high column density dusty region to the W direction of the central source can be further investigated at longer wavelengths. This region of the sky is covered by the sub-millimeter (sub-mm) observations from the APEX Telescope Large Area Survey of the Galaxy (ATLASGAL, \citealt{Schuller09}). ATLASGAL provides a high-sensitivity view of the inner Galaxy at 870~{\micron}, with an angular resolution of {19\arcsec}, enabling the study of regions associated with dense cold gas and dust, particularly in star-forming regions.
Figure~\ref{fig_images_submm} presents the near-infrared $JHK_s$ map overlaid with the 870~{\micron} contours (white curves). The sub-mm emission traces an elongated structure peaking at {$\sim$130\arcsec} to the W direction of the bright near-IR source within \iras. The near-IR source is at the same position as the ATLASGAL source identified as G339.7187--1.2041 \citep{Csengeri14}.
This entire sub-mm structure has a projected length of about {150\arcsec} and is located near a compact, bright sub-mm structure situated to the S direction, lying outside of the FOV of the map presented in Fig.~\ref{fig_images_submm}.
Notably, the bright point-like source position lies at the edge of this extended sub-mm structure, which accounts for the inhomogeneous morphology of the $K$-band continuum emission extending toward the NE direction.

We further compared the near-IR and sub-mm maps with Australia Telescope Compact Array (ATCA) radio observations at 8.6 GHz (3~cm, beam size of 2\farcs11$\times$1\farcs21, red lines) and  4.8 GHz (6~cm, 3\farcs08$\times$1\farcs76, black lines), presented by \citet{Urquhart07}. These observations are indicated as the red (3~cm) and black contours (6~cm) in Fig.~\ref{fig_images_submm}.
As typically observed in compact \hii regions, the centimeter emission traces a compact region coinciding with the peak intensity of the near-IR emission.
The 3-cm emission is more extended than the 6-cm emission, with both structures oriented toward the east of the source.

The centimeter emission exhibits a cometary morphology, consistent with the classification of \uchii regions proposed by \citet{Wood89}, indicating a parabolic ionization front.
This particular morphology is often attributed to the supersonic motion of the ionization front through the surrounding neutral gas.
However, the \brg contours (blue curves) reveal that the ionized front encompasses a much larger area than the centimeter emission, which primarily probes the peak intensity of the ionized gas surrounding the central star.

\section{The near-infrared spectroscopic analysis of IRAS 16475--4609}
\label{sec_spectroscopy}

\begin{figure*}[!ht]
    \centering
    {\includegraphics[width=\linewidth]{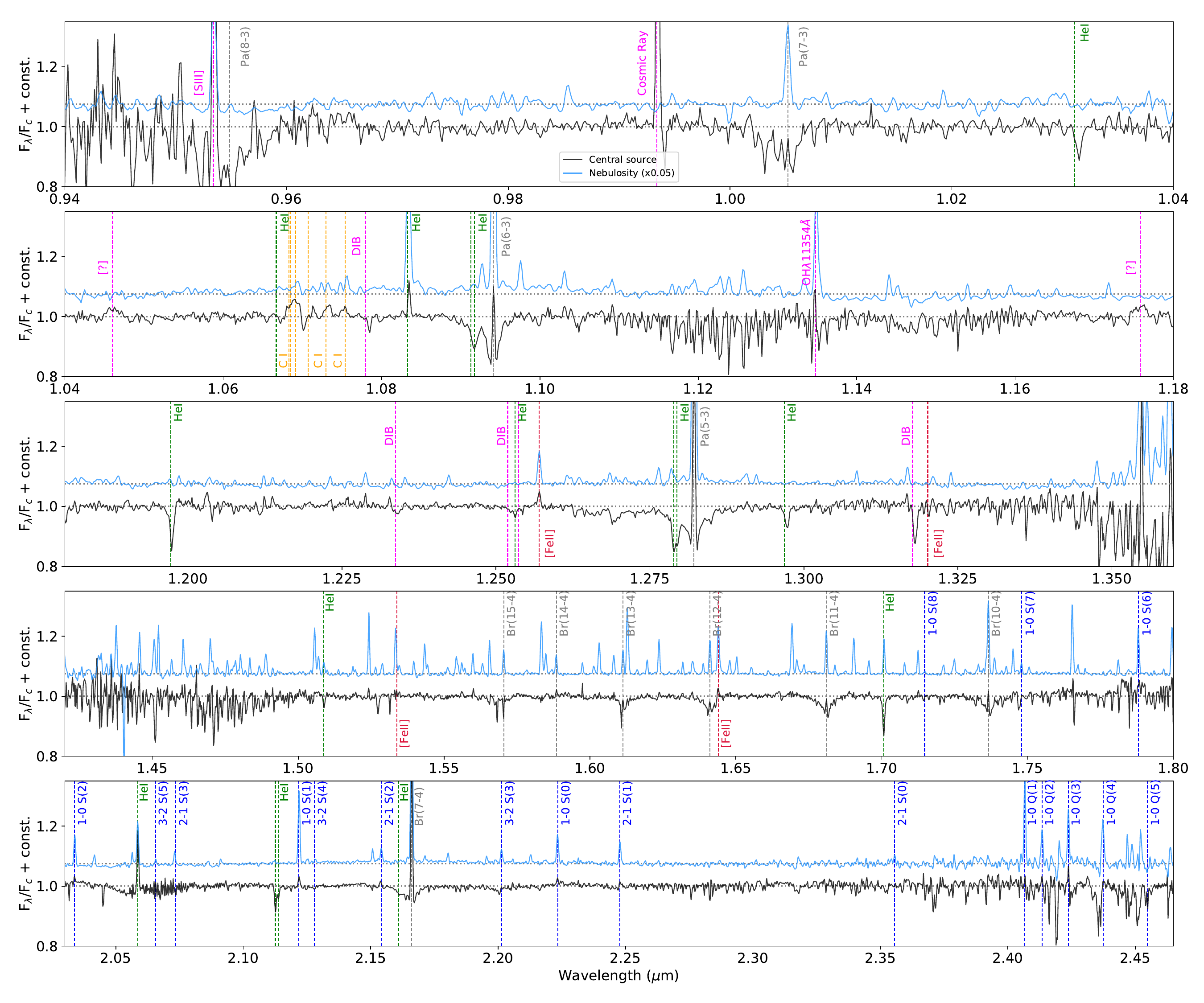}} \\[-2.0ex]
    \caption{Continuum-normalized near-IR spectrum of IRAS~16475 (black curve) and the associated nebulosity (blue). The nebular spectrum was scaled by a factor of 0.05 with respect to the central source. Vertical dashed lines indicate the wavelength of relevant spectral features: \ion{H}{1} (gray lines) and \ion{He}{1} (blue), \ion{Fe}{2} (red), \ion{C}{1} (yellow), [\ion{S}{2}], DIB and some unknown transitions (magenta).}
    \label{fig_norm_spectrum}
\end{figure*}

In this section, we present the spatially integrated analysis of the near-IR spectrum of the central source powering the \hii region and the associated nebulosity towards \iras.
The one-dimensional normalized spectrum of the point source and the extended nebular emission (scaled by a 0.05 factor) are presented in Fig.~\ref{fig_norm_spectrum}.
Strong nebular emission lines, such as \brg and \ion{He}{1} at 2.0587~{\micron}, have been suppressed to highlight the weaker lines.

The point source spectrum (black curve) exhibits relatively broad photospheric \ion{H}{1} recombination lines in absorption, typical features observed in main sequence stars.
A narrow \ion{H}{1} emission component is superimposed on these photospheric absorption features, likely arising from the surrounding nebulosity detected in the near-infrared images (Fig.~\ref{fig_images}).
The near-infrared spectrum was used to classify the spectral type of the central source in Sect.~\ref{sec_sptype}.

The nebular spectrum (blue curve) exhibits a typical spectrum of an \hii region, with the \ion{H}{1} features observed as narrow emission lines.
Additional emission features include \ion{He}{1} at 1.083 and 2.058~{\micron}, a strong forbidden [\ion{S}{3}] line at 0.9534~{\micron}, and at least four [\ion{Fe}{2}] features between 1.25 and 1.65~{\micron}.
Several molecular hydrogen (\hh) lines are also identified in the nebular spectrum.
Fainter {unmarked} emission features, particularly toward longer wavelengths, correspond to residual telluric emission lines {that were not fully removed from the data}.
The nebular spectrum is further analyzed in Sect.~\ref{sec_nebular}, with a detailed analysis of the \hh emission in Sect.~\ref{sec_h2}.

\subsection{Characterization of the central source}
\label{sec_sptype}

Fig.~\ref{fig_norm_spectrum} presents the normalized spectrum of the central source towards \iras.
We will discuss the identified spectral features across the full wavelength range covered by our observations, with an emphasis on the H- and K-band windows, which are crucial for spectral classification based on the scheme from \citet{Hanson96,Hanson05}. The equivalent widths of the key spectral lines are summarized in Table \ref{table_ews}.

Due to the non-uniform distribution of the nebular emission along the slit (see Figs.~\ref{fig_spec2d} and \ref{fig_emission1d}), the \ion{H}{1} features exhibit complex profiles. Notably, the Pa$\beta$ line at 1.28~{\micron} and \brg at 2.1661~{\micron} exhibit a combination of narrow nebular emission superimposed on broad absorption features.

We detect the Pa$\gamma$ and Pa$\delta$ transitions at 1.094 and 1.005~{\micron}, respectively, though Pa$\delta$ is blended with a strong, narrow [\ion{S}{3}] emission feature at 0.9532~{\micron}, which appears on its blue wing.

The $H$-band exhibits the higher-order Brackett series transitions up to (15-4) at 1.5705~{\micron}. Higher transitions are not detected likely due to the limited signal-to-noise ratio (SNR) of the data. Similar to \brg, these Brackett lines also exhibit narrow emission components superimposed on the broad absorption feature, indicating the complex surrounding nebular environment.

Nebular emission is likely associated with two \ion{He}{1} features: 2$^3$P-2$^3$S at 1.083 and 2$^1$P-2$^1$S 2.058~{\micron}; while several other \ion{He}{1} transitions are detected only in absorption.
The strongest \ion{He}{1} absorption features correspond to the $5^{3}D-3^{3}P$ transition at 1.1973~{\micron}, the $4^{3}P-3^{3}P$ at 1.70 ~{\micron} and the $4^{3}S-3^{3}P/4^{1}S-3^{1}P$ doublet at 2.11~\micron with no clear emission counterpart.

A few relatively weak J-band emission features were detected, in particular the \ion{C}{1} multiplet at $\lambda$10686\AA, including the $^3$P$^0_2$--$^3$D$^1$ feature at $\lambda$10675\AA, absent in the spectral atlas from \citep[][see their Table~2]{Groh07}. Two relatively weak and broad emission features with no clear identification are labeled at the following wavelengths: $\lambda$10460\AA and $\lambda$11758\AA.
The bright peak at $\lambda$9934\AA is identified as a cosmic ray residual, and a strong residual peak at $\lambda$11348\AA is attributed to poor sky subtraction within the H$_2$O telluric band between 1.11-1.16~{\micron}. 
A few diffuse interstellar band (DIBs) candidates were identified at 1.234, 1.253, 1.297 and 1.318~{\micron}, consistent with those reported by \citet{Hamano22}.

\begin{table}[!ht]
\centering
\caption{Equivalent width measurements for \ion{H}{1}, \ion{He}{1}, and DIB features identified in the stellar spectrum of \iras.}
    \label{table_ews}
\begin{tabular}{l|cccc}
\hline
\hline
Line  &   Transition & $\lambda_0$ & $\lambda_{obs}$ & EW \\
      &              & (\AA)       & (\AA)    & (\AA) \\
\hline
\ion{He}{1} &  $6^{3}S-3^{3}P$ & 10667 & 10672.61 & 0.1797 \\
\ion{He}{1} &  $2^{3}P-2^{3}S$ & 10830 & 10834.77 & -0.3835  \\
\ion{He}{1} &  $5^{3}D-3^{3}P$ & 11970 & 11973.21 & 1.1815  \\
\ion{He}{1} &  $5^{3}F-3^{3}D$ & 12789 & 12791.07 & 1.0679  \\
\ion{He}{1} &  $5^{1}D-3^{1}P$ & 12986 & 12972.56 & 0.4377 \\
\ion{He}{1} &  $4^{1}P-3^{1}S$ & 15084 & 15089.06 & 0.4331 \\
\ion{He}{1} &  $4^{3}P-3^{3}P$ & 17002 & 17006.99 & 1.2192  \\
\ion{He}{1} &  $2^{1}P-2^{1}S$ & 20580 & 20586.88 & -1.2158 \\
\ion{He}{1} &  $4^{3}S-3^{3}P$ & 21126 & 21129.98 & 1.0957  \\
\hline
\ion{H}{1} & Br(15) & 15705 & 15686.82 & 1.5483 \\
\ion{H}{1} & Br(14) & 15885 & 15822.56 & 0.4522 \\
\ion{H}{1} & Br(13) & 16114 & 16111.53 & 1.0125 \\
\ion{H}{1} & Br(12) & 16412 & 16407.19 & 1.9932 \\
\ion{H}{1} & Br(11) & 16811 & 16807.89 & 2.8048 \\
\ion{H}{1} & Br(10) & 17367 & 17366.25 & 1.5490 \\ 
\ion{H}{1} & Br(8)  & 19451 & 19450.30 & 3.5946 \\
\ion{H}{1} & Br(7)  & 21661 & 21657.67 & -0.0237 \\
\hline
\ion{H}{1} & Pa(7)  & 10052 & 10044.89 & 3.4552 \\
\ion{H}{1} & Pa(6)  & 10941 & 10933.87 & 3.1441 \\
\ion{H}{1} & Pa(5)  & 12822 & 12814.82 & 2.2834 \\
\hline
DIB        &        & 10280 & 1280.90  & 0.1582 \\
DIB        &        & 12337 & 12339.82 & 0.3285 \\
DIB        &        & 12692 & 12692.49 & 0.5933 \\
DIB        &        & 13180 & 13180.92 & 0.9562 \\
DIB        &        & 15273 & 15273.68 & 0.5283 \\
DIB        &        & 15653 & 15654.00 & 0.2438 \\
DIB        &        & 16573 & 16575.12 & 0.0551 \\
\hline
\end{tabular} \\
{{Notes:} The columns are as follows: (a) Atom, (b) transition, (c) Rest wavelength, (d) Observed wavelength, (e) equivalent width.}  
\end{table}

\subsubsection{Spectral Type Classification}

The normalized K-band spectrum of \iras reveals no clear detection of spectral features typically associated with early- and mid-O type stars, such as the \ion{C}{4} triplet, the \ion{N}{3} emission, or \ion{He}{2} absorption.
The absence of \ion{N}{3} emission implies a spectral type no earlier than O9~V, while the strong \ion{He}{1} absorption at 2.11~{\micron} suggests a spectral type earlier than B5~V.
Given the absence of ionized \ion{He}{2} lines, the spectral type classification scheme for late-O and early-B stars relies on the relative strengths of \ion{H}{1} and \ion{He}{1} features, which are less sensitive to stellar parameters than those of early to mid O-type stars \citep{Hanson96,Hanson05}. 

A visual comparison between the spectrum of \iras and the OB atlases from \citet{Hanson96,Hanson05} suggests a spectral type ranging from O9.5~V (e.g., HD~37468, which exhibits relatively weaker \ion{H}{1} lines compared to those observed in Fig.~\ref{fig_norm_spectrum}) to B2~V (e.g., HD~36166, which shows weaker \ion{He}{1} absorption at 2.11~{\micron}).

To further refine the spectral type classification, we took advantage of a complementary atlas of OB stars observed with TripleSpec (Navarete \textit{et al.}, in prep.), which extends the spectral coverage of the OB star atlases from \citet{Hanson96,Hanson05} to a broader wavelength range of 0.94–2.47~{\micron}, providing data with relatively high signal-to-noise ratios (SNR).
The TripleSpec atlas encompasses luminosity class V stars from early O to early A-type, including all equatorial and southern targets from \citet{Hanson96,Hanson05}, along with additional well-known OBA stars observable from the SOAR telescope (Decl.~$<$~30$^\circ$). This expanded coverage allows for a more precise comparison across a wider spectral range, matching the same spectral resolution of \iras observations (R$\sim$3500).

In Fig.~\ref{fig_spec_comparisonOB} (Appendix~\ref{appendix_OB}), we present four key spectral regions that highlight the primary features used for the spectral type classification of \iras, comparing its normalized spectrum with six OB standards spanning the O9.5-B2~V range.
A more detailed analysis of the \ion{H}{1} features narrows the spectral type of the central source to the O9.5-B0.7~V range, refining the broader classification initially derived from visual inspection of the OB star atlases by \citet{Hanson96,Hanson05}.

\begin{figure*}[!ht]
    \centering
    {\includegraphics[width=\linewidth]{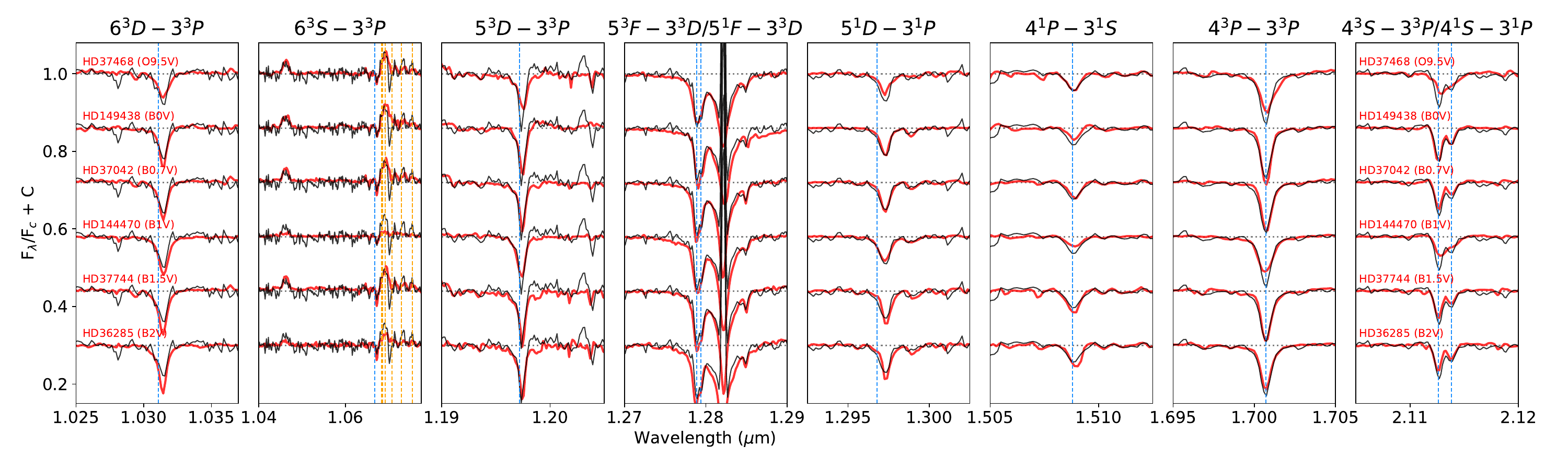}} \\[-1.0ex]
    \caption{Zoom around \ion{He}{1} lines {in the} continuum-normalized {near-IR} spectrum of \iras.
    The red curves are a sample of representative OB-type stars observed with TripleSpec (Navarete \textit{et al.}, in prep.), from top to bottom: HD37468(O9.5~V), HD149438 (B0~V), HD37042 (B0.7~V), HD144470 (B1~V), HD37744 (B1.5~V), and HD36285 (B2~V). The vertical dashed lines indicate \ion{He}{1} (blue), and \ion{C}{1} (yellow) lines used for classification.}
    \label{fig_spec_comparisonOB_helium}
\end{figure*}

We further compared the emission-free \ion{He}{1} line profiles from the \iras spectrum with those of the OB standards observed with TripleSpec to check if the spectral type classification could be improved. Figure~\ref{fig_spec_comparisonOB_helium} presents eight spectral windows containing \ion{He}{1} lines that serve this purpose.
In general, most of the \ion{He}{1} absorption features match the profiles observed in B0 and B0.7~V stars, including the \ion{He}{1} absorption at 1.297~{\micron}, which is strongly affected by the blue wing of the Pa$\beta$ line.
In addition, the strength of the \ion{C}{1} multiplet around 10686~{\AA} further supports a classification of the central source as an early-B star, no later than B0.7~V.

\subsection{Analysis of the extended nebular emission}
\label{sec_nebular}

We took advantage of the spatial information provided by the TripleSpec spectroscopic observations to further investigate the extended emission detected in the narrow-band images of \iras (Fig.~\ref{fig_images}). 

Spectroimages, which display spectral dispersion along the slit direction, were extracted for selected emission lines using the \texttt{IDL} \texttt{tspec\_extract\_lines} procedure (see Sect.~\ref{sec_spectroimages_def}).
This method applies an outlier-resistant polynomial fit to model the continuum, which is then subtracted from the data. The emission lines are modeled with Gaussian profiles using the \texttt{MPFITPEAK} routine.

\begin{figure}[!ht]
    \centering
    {\includegraphics[width=\linewidth]{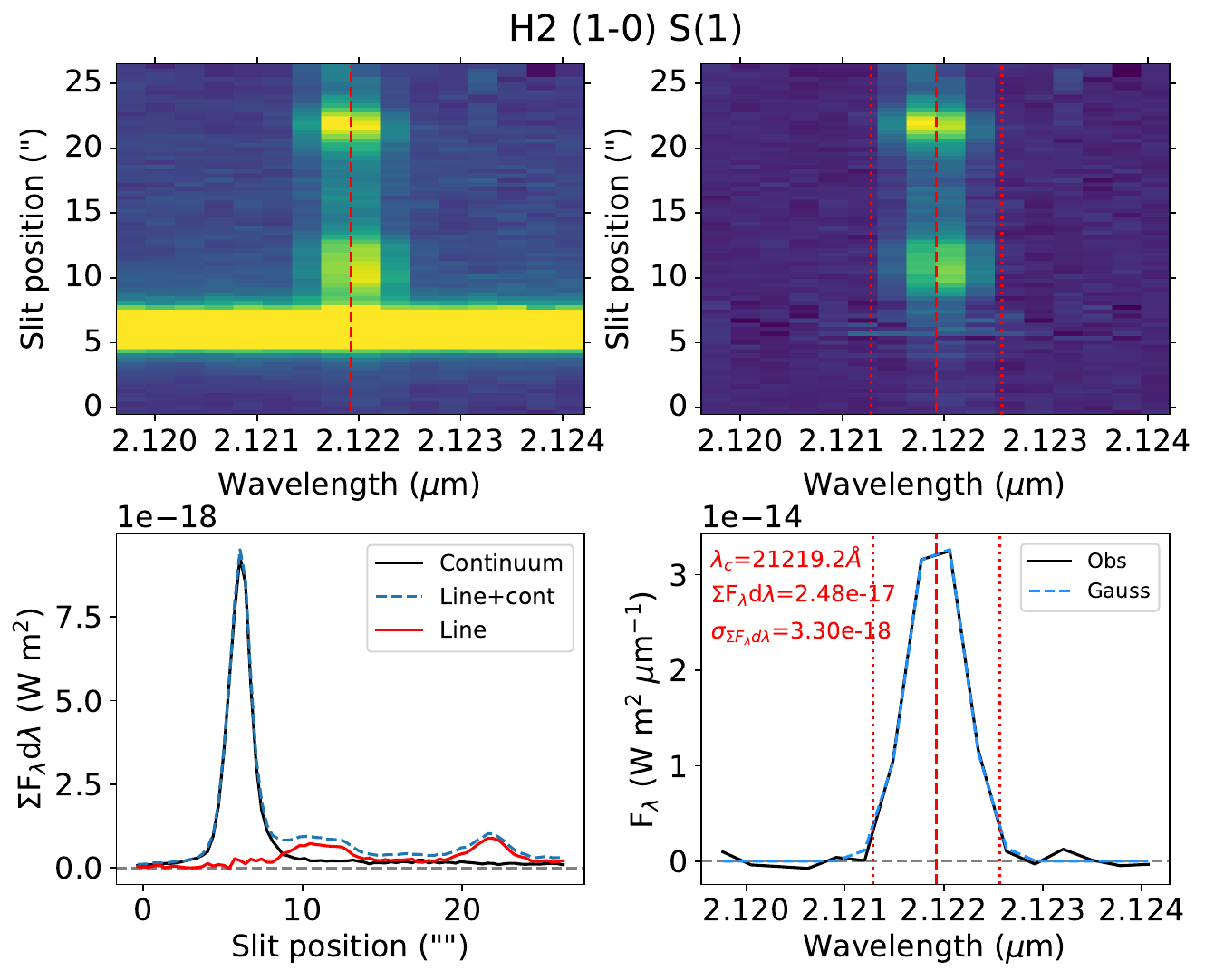}} \\[0.0ex]
    \caption{Spectro image of the \hh~(1-0)~S(1) transition at 2.1218~{\micron} towards \iras.
    Top panels: Spectroimages of the total flux (line+continuum, left) and the continuum-subtracted emission (right). The vertical dashed red line marks the center of the profile, while the vertical dotted red lines indicate the spectral range used to integrate the total flux.
    Bottom left: Spatial distribution along the slit for the (a) total flux (line+continuum, dashed blue curve), (b) continuum-subtracted line emission (solid red curve), and (c) continuum emission (solid black).
    Bottom right: Continuum-subtracted integrated spectrum. The vertical lines are the same as defined in the top right panel. The Gaussian fit to the emission line is shown by the dashed blue curve. The central wavelength, the integrated flux and its error are indicated in the top left corner.
    }
    \label{fig_specimg}
\end{figure}

The total flux ($\Sigma F_\lambda d\lambda$, in W~cm$^{-2}$ units) is integrated within a 90\% confidence interval based on the fitted model.
Figure~\ref{fig_specimg} illustrates the result for the \hh~(1-0)~S(1) transition at 2.1218~{\micron}.

\begin{figure}[!ht]
    \centering
    {\includegraphics[width=\linewidth]{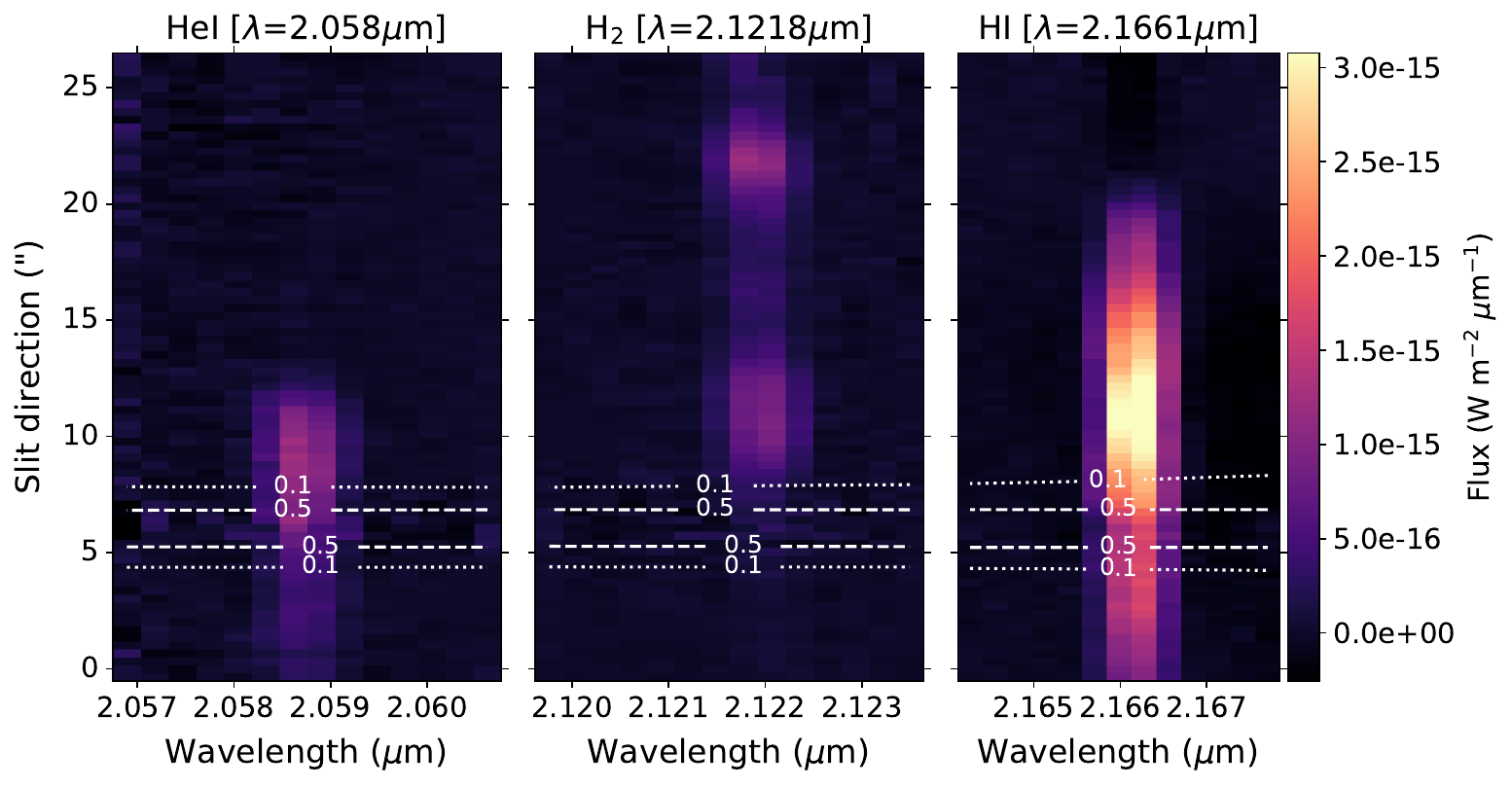}} \\[0.0ex]
    \caption{Spectroimages of the \ion{He}{1} ($\lambda$~=~2.0587~{\micron}, left panel), \hh~(1--0)~S(1) ($\lambda$~=~2.1218~{\micron}, middle) and \brg  ($\lambda$~=~2.1661~{\micron}, right).
    The color bar to the right indicates the flux intensity for all three features.
    White contours represent the continuum emission, with the dashed and the solid lines at 10 and 50\% of the peak intensity, respectively, marking the position of the point-like source.}
    \label{fig_spec2d}
\end{figure}

Figure~\ref{fig_spec2d} presents the continuum-subtracted spectroimages for the \ion{He}{1} ($\lambda$~=~2.0578~{\micron}, the \hh~(1--0)~S(1) ($\lambda$~=~2.1218~{\micron}), and the \brg ($\lambda$~=~2.1661~{\micron}) features. 
The spatial axis (y-axis) reveals distinct morphologies for each transition, suggesting that they trace different phases of the extended emission associated with \iras. No significant clear velocity gradient is observed along the spectral direction (x-axis) for any of these features.

\begin{figure}[!ht]
    \centering
    {\includegraphics[width=\linewidth]{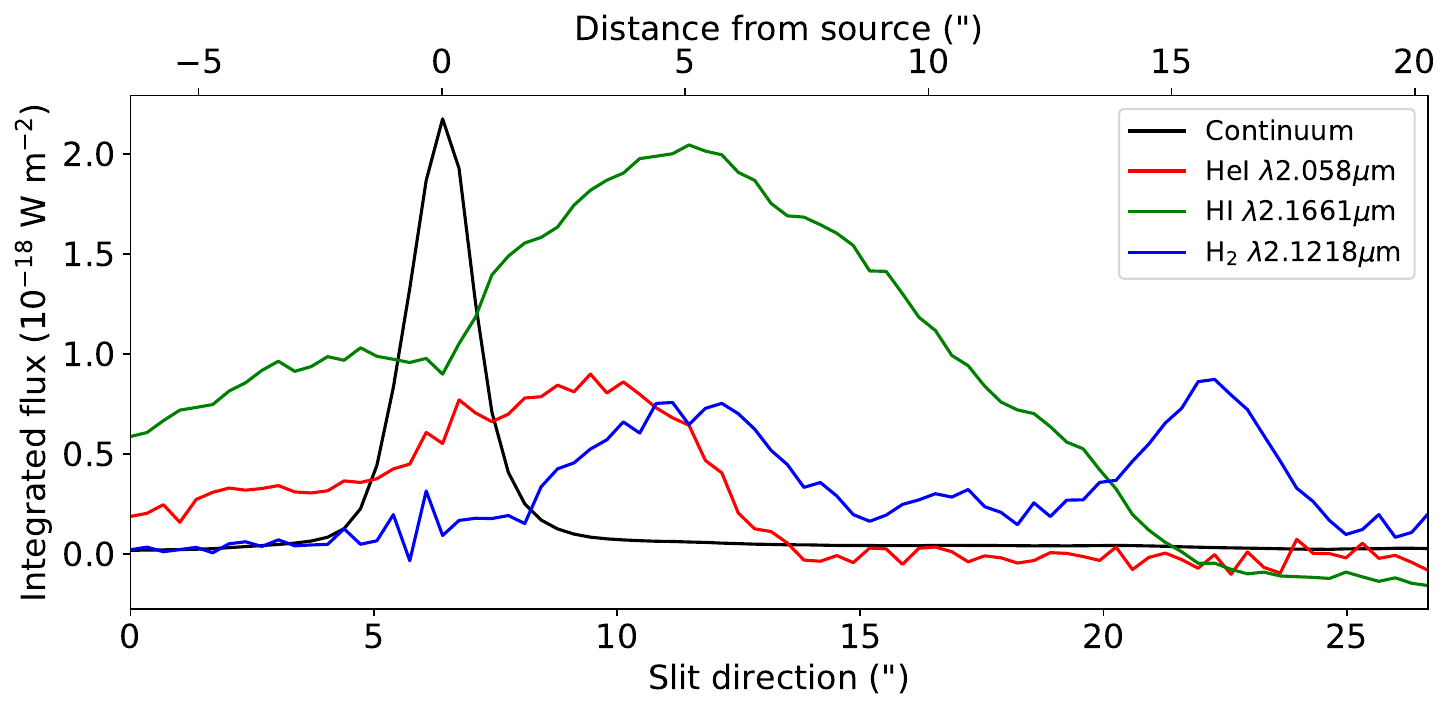}} \\[-1.0ex]
    \caption{Integrated flux of the extended emission along the slit. The \brg, \hh. and \ion{He}{1} emission from Fig.~\ref{fig_spec2d} are represented by the green, blue, and red curves, respectively. The black curve indicates the position of the point-like source. The top x-axis shows the slit direction in arcsecond units centered on the location of the point-like source.}
    \label{fig_emission1d}
\end{figure}

To better visualize the flux distribution along the slit, Fig.~\ref{fig_emission1d} shows the integrated flux of each spectral line as a function of the slit position, alongside the K-band continuum emission (black curve). 

The continuum emission peaks sharply at the position of the central source, indicating a minimal contribution from extended continuum emission.
The \brg emission (green curve) traces extended ionized gas, peaking $\approx$5{\arcsec} from the central source, and gradually decreases out to $\sim$16\arcsec. This distribution is consistent with a wide-angle conical outflow of ionized gas, with the flux increasing with distance from the star due to the expanding ionization front, reaching a maximum where ionization is most efficient, and then decreasing as fewer neutral atoms can still be excited by the UV radiation from the central star.

The \ion{He}{1} emission (in red) exhibits a similar, though more compact, spatial profile, extending to $\sim$6{\arcsec} from the central star and peaking closer, at $\sim$3{\arcsec}. This suggests that the He excitation radius is smaller than that of H, consistent with the fact that He requires higher energies to be excited.

The molecular hydrogen line (in blue) displays a double-peak profile, with emission maxima at {$\sim$5\arcsec} and {$\sim$16\arcsec} from the central source. The inner peak coincides with the outer boundary of the \ion{He}{1} distribution and the peak of the \brg emission, while the outer peak encircles the \hii region in a shell-like structure, likely tracing the photodissociation region (PDR).
Between these peaks, the \hh emission persists at a lower level, suggesting a gradual transition zone where the molecular gas is partially shielded from direct stellar radiation, but still influenced by the expanding ionization front. This non-zero emission could indicate a clumpy or filamentary distribution of molecular material, allowing for survival even within regions closer to the ionization boundary.

\subsubsection{Molecular \texorpdfstring{H$_2$}{H2} analysis}
\label{sec_h2}

We constructed spectroimages for selected \hh lines to extract their integrated flux.
Table~\ref{table_h2} lists the fluxes and the physical properties of K-band \hh transitions detected above the 1-$\sigma$ level. 
The integrated fluxes were obtained from the corresponding spectroimages for each transition (e.g., Fig.~\ref{fig_spec2d}). Using the continuum-subtracted fluxes of the individual \hh lines, we constructed a ro-vibrational Boltzmann diagram of the \hh molecule to derive the physical parameters of the molecular gas, assuming local thermodynamic equilibrium (LTE).

\setlength{\tabcolsep}{3pt}
\begin{table}[!ht]
    \centering
    \caption{K-band \hh fluxes extracted from spectroimages from the near-IR spectroscopic observations towards \iras.}
    \label{table_h2}
    \begin{tabular}{crrrrrr}
\hline\hline
Transition & \multicolumn{1}{c}{Flux} & \multicolumn{1}{c}{$\sigma_{F}$} & \multicolumn{1}{c}{$\lambda$} & \multicolumn{1}{c}{$g$} & \multicolumn{1}{c}{$E_{upper}$} & \multicolumn{1}{c}{$A$} \\
           & \multicolumn{2}{c}{($10^{-18}$ W~m$^{-2}$)} & \multicolumn{1}{c}{(\micron)} &  & \multicolumn{1}{c}{(K)} & \multicolumn{1}{c}{(10$^{-7}$~s$^{_1}$)} \\
\hline
2-1~S(5) &	9.24	&	0.91	&	1.9449	&	45	&	15763	&	5.05	\\
1-0~S(3) &	32.30	&	0.96	&	1.9576	&	33	&	 8365	&	4.21	\\
1-0~S(2) &	11.00	&	1.02	&	2.0338	&	 9	&	 7584	&	3.98	\\
3-2~S(5) &	 1.94	&	0.97	&	2.0656	&	45	&	20856	&	4.50	\\
1-0~S(1) &	24.20	&	0.94	&	2.1218	&	21	&	 6956	&	3.47	\\
3-2~S(4) &	 1.18	&	0.93	&	2.1280	&	13	&	19912	&	5.22	\\
2-1~S(2) &	 6.22	&	1.01	&	2.1542	&	 9	&	13150	&	5.60	\\
3-2~S(3) &	 2.59	&	0.85	&	2.2014	&	33	&	19086	&	5.63	\\
1-0~S(0) &	10.10	&	0.96	&	2.2235	&	 5	&	 6471	&	2.53	\\
2-1~S(1) &	 7.67	&	0.92	&	2.2477	&	21	&	12550	&	4.98	\\
3-2~S(2) &	 2.08	&	0.87	&	2.2870	&	 9	&	18386	&	5.63	\\
3-2~S(1) &	 3.91	&	0.93	&	2.3864	&	21	&	17818	&	5.14	\\
1-0~Q(1) &	36.30	&	0.75	&	2.4066	&	 9	&	 6149	&	4.29	\\
1-0~Q(2) &	12.20	&	0.70	&	2.4134	&	 5	&	 6471	&	3.03	\\
1-0~Q(3) &	25.00	&	0.69	&	2.4237	&	21	&	 6956	&	2.78	\\
1-0~Q(4) &	6.58	&	0.84	&	2.4375	&	 9	&	 7586	&	2.65	\\
\hline
    \end{tabular} \\
{Notes:} the energy levels are from \citet{Dabrowski84}, and the Einstein coefficients are from \citet{Turner77}.
\end{table}
\setlength{\tabcolsep}{6pt}

Assuming a Boltzmann distribution and optically thin emission, the observed \hh fluxes $F_J$ follow the relation:
\begin{equation}
\frac{4 \pi F_J}{\Omega h \nu_J g_J A_J} = \frac{N_J}{g_J} = \frac{N_{tot}}{Q_{rot}} \exp\left({-E_J/k_B T_{exc}}\right)
\label{eq_boltzmann}
\end{equation}
\noindent where $h$ is the Planck constant,
$k_B$ is the Boltzmann constant (in erg~K$^{-1}$),
$\Omega$ is the emitting area solid angle (in sr),
$\nu_J$ is the frequency (in Hz), 
$g_{J}$ is the statistical weight, 
$A_{J}$ is the Einstein coefficient (in s$^{-1}$),
$N_{J}$ is the upper-level column density of the \hh transition (in cm$^{-2}$),
$N_{tot}$ is the total column density of the \hh molecule (in cm$^{-2}$),
$Q_{rot}$ is the partition function, 
$E_J$ is the upper-level transition energy (in erg), and 
$T_{rot}$ is the rotational temperature (in K).

The total column density and temperature of the gas are estimated through the linear fit defined as:
\begin{equation}
    \ln\left(\frac{N_J}{g_J}\right) = \ln\left(\frac{N_{tot}}{Q_{rot}}\right) -\frac{E_J}{k_B T_{rot}} = \beta + \alpha E_J
\label{eq_boltzmann_linear}
\end{equation}

\begin{figure}[!ht]
    \centering
    \includegraphics[width=\linewidth]{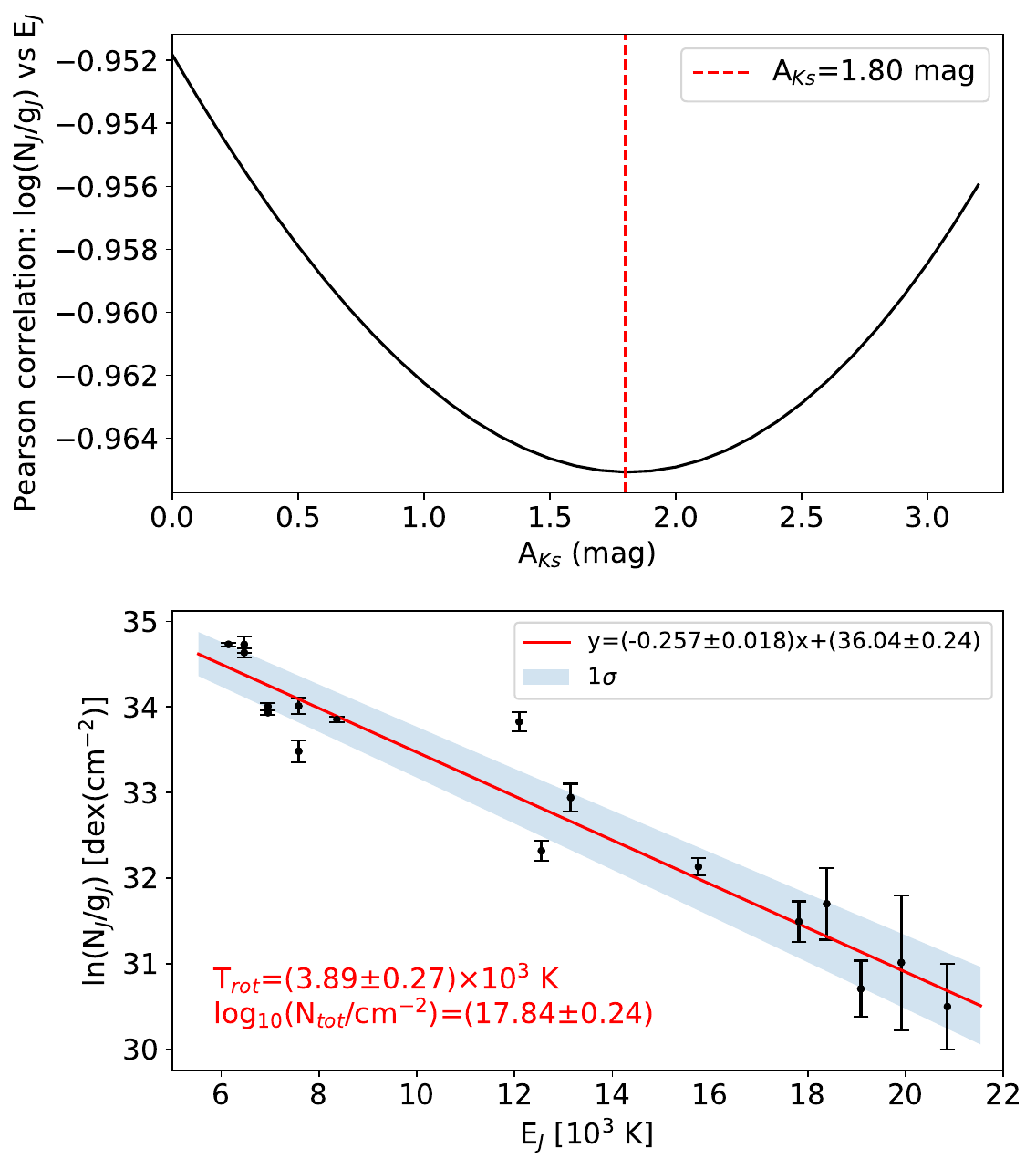} \\[-2.0ex]
    \caption{Analysis of the molecular hydrogen emission associated with the nebular component of \iras. Top panel: Pearson correlation between the logarithm of the upper-level column density as a function of the upper-level energy of the \hh transitions by varying the \ak values from 0 to 3.3~mag in steps of 0.1~mag. The best \ak value is indicated by the vertical dashed red line, corresponding to the value that maximizes the correlation.
    Bottom panel: Boltzmann diagram of the \hh emission after correcting for reddening effects. The solid red line indicates the best fit to the data (values in the legend), and the shaded area shows the 1$\sigma$ region of the fit. The corresponding \trot and \ntot values from the best fit are indicated as red text.}
    \label{fig_h2diagram}
\end{figure}

Eqs.~(\ref{eq_boltzmann}) and (\ref{eq_boltzmann_linear}) require reddening-corrected fluxes and the latter can also be used to infer the K-band extinction (\ak). By following the procedure outlined by \citet{Caratti15}, one can vary \ak to identify the value that maximizes the correlation between $\ln\left(\frac{N_J}{g_J}\right)$ and $E_J$.
The top panel of Fig.~\ref{fig_h2diagram} shows the Pearson correlation coefficient as a function of the $K$-band reddening, indicating that the strongest correlation occurs for \ak~=~1.80$\pm$0.1~mag (corresponding to  $A_{V}$~=~26.9$\pm$1.5~mag assuming the $A_{\rm V}/A_{\rm K}$ ratio of 14.95 from \citealt{Damineli16}).

The bottom panel of Fig.~\ref{fig_h2diagram} presents the Boltzmann diagram of the \hh emission, corrected for extinction. The best fit line yields a slope of $\alpha$~=~$-$0.257$\pm$0.018, corresponding to a \trot value of (3.89$\pm$0.27)$\times$10$^3$~K.
The linear coefficient corresponds to a total \hh column density of $\log_{10}$(N$_{tot}$/cm$^{-2}$)~=~17.8$\pm$0.2~dex. 

\begin{figure}[!ht]
    \centering
    {\includegraphics[width=\linewidth]{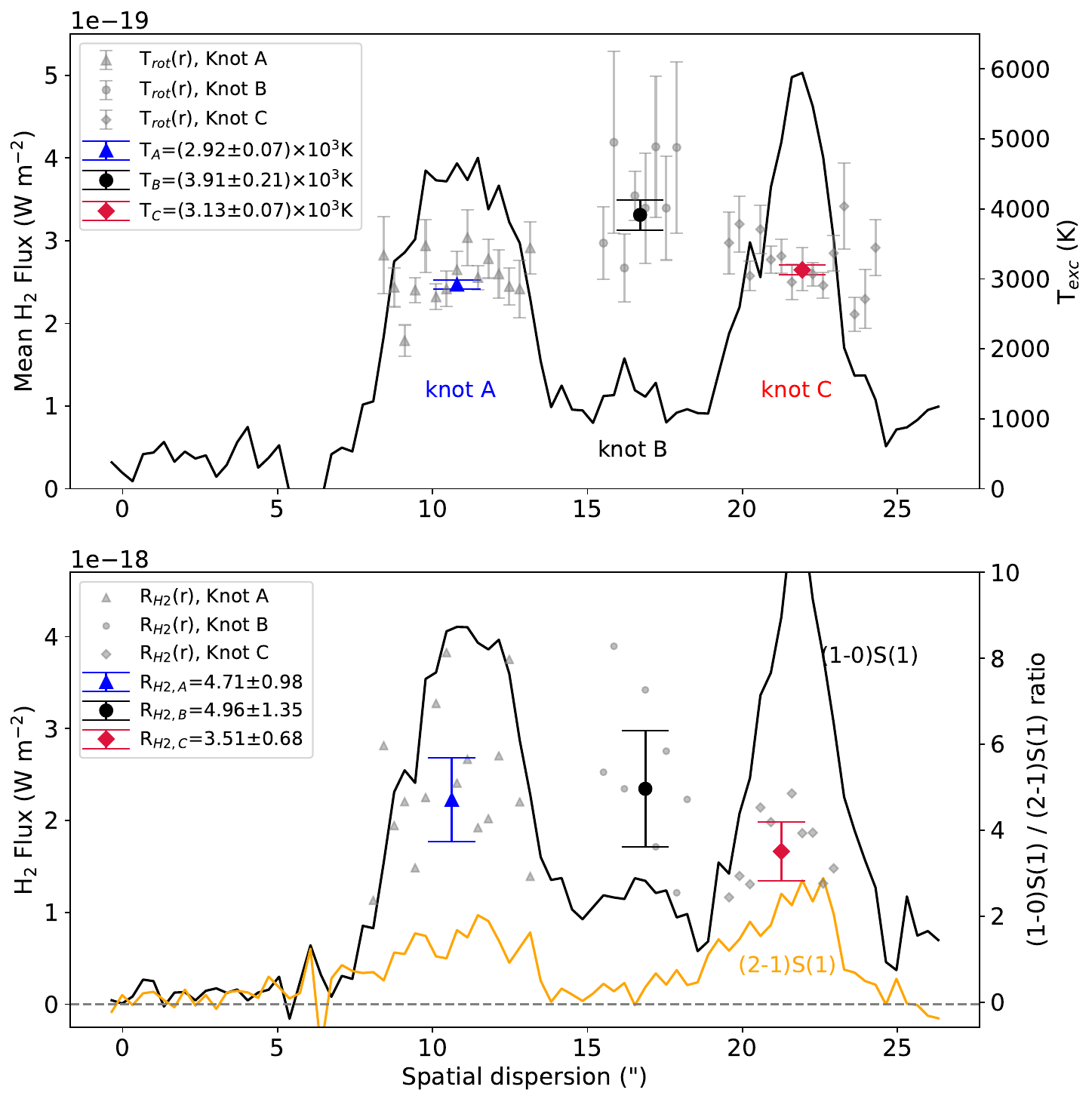}} \\[-2.0ex]
    \caption{Top panel: Spatial distribution of the mean flux of 11 \hh transitions along the slit direction (black curve). The gray points correspond to the \trot of the gas evaluated at each spaxel. The colored points correspond to the weighted mean \trot values for each of the three labeled \hh knots: A (blue triangle), B (black circle) and C (red diamond).
    Bottom: Spatial distribution of the (1--0)~S(1) and (2--1)~S(1) transitions of the molecular hydrogen (blue and orange curves, respectively) corrected for extinction assuming \ak~=~1.80~mag (Fig.~\ref{fig_h2diagram}), and the line ratio measurements (grey points). The colored points correspond to the weighted mean \trot values for each of the three labeled \hh knots: A (blue triangle), B (black circle), and C (red diamond)}
    \label{fig_h2diagram_p2p}
\end{figure}

Figure~\ref{fig_h2diagram_p2p} presents the spatial distribution of the \hh emission along the slit. The black curve represents the mean flux of the transitions listed in Table~\ref{table_h2}. The spatial profile reveals three distinct emission knots, {labeled as} A, B and C.
Taking advantage of the spatial information in the spectroscopic data and fixing the \ak value at 1.80~mag, we evaluated the \trot of the gas for each spatial pixel (spaxel) associated with the identified \hh knots.
A weighted mean \trot value was then calculated for each knot (represented by the larger circles with error bars in the plot). Knot B, the weaker of the three, shows a slightly higher rotational temperature of \trot~=~(3.91$\pm$0.21)$\times$10$^{3}$~K when compared to the brighter knots A and C (\trot~$\sim$3$\times$10$^{3}$~K).

The line ratios of \hh transitions are powerful diagnostics in differentiating between excitation in shocks and {PDRs} formed by UV photons. For instance, the origin of the \hh emission can be inferred using the ratio of \hh transitions such as (1--0)~S(1)/(2--1)~S(1). From the fluxes listed in Table~\ref{table_h2} we derived a (1--0)~S(1)/(2--1)~S(1) ratio of 3.6~$\pm$~0.5 after correcting for reddening effects. 
As expected in a \hii region, this ratio is closer to the predicted value of 1.8 for radiative excitation by UV photons from the protostar, rather than the theoretical value of $\sim$10 expected for thermal excitation through shocks \citep{Black76}.
As indicated in Table~\ref{table_h2}, we do detect a few $v$=3-2 transitions, strongly suggesting that pure fluorescence may indeed occur, indicating excitation in a PDR rather than in shocks.

The bottom panel of Fig.~\ref{fig_h2diagram_p2p} shows the integrated extinction-corrected fluxes for the two \hh transitions (black: (1--0)~S(1), orange: (2--1)~S(1)) along the spatial direction of the slit. The line ratio was calculated at each spaxel with fluxes above a 1-$\sigma$ threshold for both transitions. The median ratio for each knot is indicated by the blue, black, and red points with associated error bars. The median ratios range from 3.5 to 5.0, consistent with the scenario of radiative excitation through UV photons, as expected for a \hii region.

\section{Distance and Ionizing photon budget from \texorpdfstring{\iras}{IRAS 16475-4609}}
\label{sec_distance}

The accurate determination of the distance to \iras is crucial for understanding its physical properties and placing it within the broader context of star-forming regions and the Galaxy structure.
In this section, we present three independent methods for deriving the distance to \iras. 
First, we derive a spectrophotometric distance based on the spectral classification (Sect.~\ref{sec_sptype}) and near-IR photometry.
Next, we present an independent distance estimate based on the optical counterpart from Gaia Data Release 3.
Finally, we use archival radio recombination line measurements to obtain the kinematic distance, providing a third, independent determination of the distance to \iras.

With the distance to \iras, we then evaluate the ionizing photon flux and the properties of the associated ionized gas.

\subsection{The spectrophotometric distance to \texorpdfstring{\iras}{IRAS 16475-4609}}
\label{sec_spectrophotometric_distance}

The spectrophotometric distance to \iras was determined using the following approach. 
The visual absolute magnitude (M$_V$) for B0 and B0.5~V stars were obtained from \citet{Vacca96}, and their intrinsic V$-$K colors from \citet{Koornneef83}. 
To ensure consistency with different photometric systems, we applied the transformations from \citet{Carpenter01} to convert the photometry to the 2MASS system.

Photometric data for \iras were extracted from the 2MASS catalog \citet{Skrutskie06}, yielding the following values: m$_{Ks}$~=~10.255$\pm$0.034, J$-$H~=~0.925$\pm$0.040, J$-$K$_s$~=~1.490$\pm$0.043, H$-$K$_s$~=~0.565$\pm$0.045~mag. 
To account for extinction,  the $K_s$ band extinction (A$_{Ks}$) was obtained by using the near-infrared color excess and applying Eqs.~(15)-(17) from \citet{Damineli16}.
This method allowed us to correct for reddening effects and accurately derive the extinction-corrected magnitude, a critical step for determining the distance to the \hii region.

The distance modulus ($\mu=m_{Ks}-M_{Ks}-A_{Ks}$) and the corresponding distance ($d$) for each Spectral Type are reported in Table~\ref{table_spectro_distances}, along with the relevant photometric properties.
Given the uncertainty in the precise determination of the subspectral type, we computed the spectrophotometric distance for both B0 and B0.5~V subclasses individually, using their respective absolute magnitudes. We also computed the average distance by considering the mean absolute magnitude for the B0-B0.5 range. Based on this analysis, the mean distance to \iras is 3.6$\pm$1.6~kpc. 

\begin{table}[!ht]
    \centering
    \caption{Photometric information and spectrophotometric distance determination towards \iras.}
    \label{table_spectro_distances}
    \begin{tabular}{c|ccc}
    \hline\hline
 Spectral Type  & B0~V          & B0.5~V        & B0-B0.5~V   \\
    \hline
M$_{Ks}$ (mag) & --3.31$\pm$0.67 & --3.26$\pm$0.67 & --3.28$\pm$0.95 \\
\ak (mag)      &   0.75$\pm$0.07 &   0.74$\pm$0.07 &   0.75$\pm$0.10 \\
$\mu$ (mag)    &  12.81$\pm$0.67 &  12.77$\pm$0.67 &  12.79$\pm$0.95 \\
$d$ (kpc)      &    3.7$\pm$1.2  &    3.6$\pm$1.1  &     3.6$\pm$1.6 \\
\hline
    \end{tabular}
\end{table}
\setlength{\tabcolsep}{6pt}

\subsection{Photogeometric distance from Gaia DR3}
\label{sec_gaia_distance}

We searched for {the} optical counterpart of \iras using Gaia Data Release 3 (Gaia DR3; \citealt{Gaia16,Gaia23}), identifying two Gaia sources within 0\farcs84 of each other.
The brighter source, Gaia~DR3~5940120644325949824, has a $G$-band magnitude of 16.6 with reliable astrometric and photometric measurements.
The fainter source, Gaia~DR3~5940120644333998464 ($G$~=~17.7~mag), has limited information in the Gaia archive and is detected only in the $G$-band.
The results presented below consider only the brightest Gaia~DR3 source with good photometric and astrometric measurements. 

\begin{table}[!ht]
    \caption{Photometric and astrometric information of the Gaia DR3 counterpart associated with \iras.}
    \label{table_gaiadr3}
    \centering
    \begin{tabular}{c|c}
\hline
\hline
Gaia ID     &  5940120644325949824   \\
\hline
G (mag)     & 16.6332$\pm$0.0039     \\
BP-RP (mag) & 3.756$\pm$0.030        \\
A$_G$ (mag) & 5.94$^{+0.38}_{-0.25}$ \\
\hline
pmra (mas~yr$^{-1}$)  & --0.19$\pm$0.14 \\
pmdec (mas~yr$^{-1}$) & --2.25$\pm$0.11 \\
parallax (mas)        & 0.22$\pm$0.12   \\
RUWE                  & 1.14            \\
d$_{geo}$ (kpc)     & 3.68$_{-1.07}^{+1.59}$ \\
d$_{photgeo}$ (kpc) & 3.50$_{-0.89}^{+1.33}$ \\
\hline
    \end{tabular}
\end{table}

The photogeometric distance reported for Gaia~DR3~5940120644325949824 corresponds to $d_{GDR3}$~=~3.50$^{+1.33}_{-0.89}$~kpc, with a re-normalised unit weight error (RUWE) below 1.4, (RUWE~=~1.14), {indicating} a good astrometric solution \citep{Lindegren18}.
This distance corresponds to a distance modulus of $\mu_{GDR3}$~=~12.72$^{+0.70}_{-0.72}$~mag, {in close agreement} with the spectrophotometric distance derived in Sect.~\ref{sec_spectrophotometric_distance}.

\subsection{Kinematic distance from Radio Recombination Lines}
\label{sec_rrl_distance}

The \emph{Southern \hii Region Discovery Survey} \citep[SHRDS,][]{Wenger2021} reported a centimeter source (designated as SHRDS925, RA~=~16:51:15.0, Dec.~=~--46:14:23.3) coincident with the near-IR source in \iras. The radio source is associated with both continuum and hydrogen recombination line emission.

The hydrogen recombination lines (H88–H112) exhibit a mean peak velocity at \vlsr~=~$-$33.2$\pm$0.2~km~s$^{-1}$ \citep[Table~7][]{Wenger2021}.
We used the \vlsr information to evaluate the kinematic distance to \iras, using the Parallax-Based Distance Calculator\footnote{\url{http://bessel.vlbi-astrometry.org/}} \citep{Reid16, Reid19}. To further constrain the distance estimate, we incorporated the proper motions of the Gaia~DR3 counterpart (from Table~\ref{table_gaiadr3}). The uncertainties in both proper motion and \vlsr were propagated in the calculation, yielding a kinematic distance of 3.47$\pm$1.26~kpc.

\subsection{Ionizing Photon Budget and Spectral Type}
\label{sec_ionizing_source}

With the distance to \iras established in the previous subsections, we now evaluate the ionizing photon flux and the properties of the associated ionized gas.

Under the assumption that the \hii region is in photoionization equilibrium and optically thin, we estimated the Lyman continuum photon flux ($N_{\text{Ly}}$, in photons\,s$^{-1}$) using the integrated radio flux densities reported by {\citet[][Table~5]{Wenger2021}, and the modified version of Equation~(7) from \citet{Carpenter90}:}
\begin{equation}
    \frac{N_{Ly}}{\text{photon~s}^{-1}} = 7.7 \times 10^{43} \left(\frac{S_{int,\nu}}{\text{mJy}}\right) \left(\frac{d}{\text{kpc}}\right)^2 \left(\frac{\nu}{\text{GHz}}\right)^{0.1}
\end{equation}
\noindent where S$_{int,\nu}$ is the integrated radio flux density at frequency $\nu$ (in mJy), $d$ is the distance (3.51$\pm$0.91~kpc, Table~\ref{table_distances}) and $\nu$ is the frequency of the observation (in GHz).
{We computed a Lyman continuum photon flux of  $N_{\text{Ly}}$~=~47.36$\pm$0.06~dex, based on a weighted mean of the individual SHRDS radio continuum measurements with frequencies ranging from 4.9 to 9.2~GHz.}

{We estimated the electron temperature (T$_e$, in K) using the radio recombination line (RRL) and the nearby continuum flux density measurements from SHRDS \citet[][Table~7]{Wenger2021}, assuming optically thin emission in LTE \citep{Shi10}:}
\begin{equation}
\tiny
    \frac{T_e}{K} = \left[
    \left( \frac{6985}{\alpha(\nu,T_e)} \right)
    \left( \frac{\nu}{\text{GHz}} \right)^{1.1}
    \left( \frac{\Delta V_{l}}{\text{km s}^{-1}} \right)^{-1}
    \left( \frac{S_{c}}{S_{l}} \right) 
    \left( 1+\frac{N(He)}{N(H)} \right)^{-1}
    \right]^{0.87}
\end{equation}
\noindent where the dimensionless factor $\alpha(\nu,T_e)$ is the order of unity \citep{Mezger67}, $\Delta V_{l}$ corresponds to the width of the line emission (in $\text{km s}^{-1}$), $S_{c}/S_{l}$ is the ratio between the peak flux of the line and the nearby continuum, and {$N(\textrm{He})/N(\textrm{H})$~=~0.068$\pm$0.023 \citep{Wenger2013}.
Based on 19 RRL transitions (in the H88-H112 range) reported by \citet{Wenger2021}, we derived a weighted mean electron temperature of $T_{\text{e}}$~=~(5.44$\pm$0.21)$\times$10$^{3}$~K. This result is consistent for all the RRL transitions available from SHRDS, indicating robust physical conditions within the ionized gas.}

\section{Discussion and Conclusions}
\label{sec_discussion}

The multi-wavelength imaging analysis of \iras reveals a complex and dynamic environment around a massive star powering a young \hii region at the edge of a molecular clump. This scenario is well aligned with the blister/champagne flow model described by \citet{Israel78} and later refined by \citet{TenorioTagle79}. In this model, an exciting star forms near the edge of a dense molecular cloud, causing ionization fronts to gradually advance into the cloud. This leads to the formation of a cavity filled with ionized gas, from which the gas flows away into space. Over time, a pressure gradient develops between the ionized gas and the neutral cloud, driving a shock that spreads the ionized material outward, creating an observable ionized nebula. Such a scenario aligns with the morphology observed in the \iras source, where the ionized region appears to expand into a conical cavity within the surrounding molecular material.

The near-infrared maps from Fig.~\ref{fig_images} exhibit an extended nebulosity with a distinct conical morphology, suggesting a cavity created by the ionizing radiation of an early B-type star. This cavity is oriented towards the NE direction at the edge of a dense molecular clump, traced by the sub-mm 870~{\micron} map from the ATLASGAL survey (see Fig.~\ref{fig_images_submm}).

The narrow-band near-IR images reveal different phases of the nebulosity: \brg and forbidden [\ion{Fe}{2}] emission trace ionized gas with a spherical morphology close to the central source, while the \hh transition at 2.12~{\micron} highlights the excited molecular gas component at larger distances.
As expected for a young \hii region, the \brg emission is compact and coincides with the position of the radio centimeter emission from ATCA, while the \hh emission is more extended, originating in the surrounding molecular gas \citep{Hanson02}. 

This scenario is supported by Figs.~\ref{fig_images} and \ref{fig_images_submm}, exhibiting the spatial distribution of these components.
The \hh emission probes the photodissociation region (PDR) at the interface between the ionized and neutral gas in the outer boundary of the extended $K$-band emission in \iras. This pattern is consistent with UV-pumped molecular gas at the edges of a dense molecular clump, a characteristic structure often observed in young \hii regions. The conical shape of the nebula to the northeast strongly suggests that the ionizing radiation has carved out an outflow cavity in this direction.
The coupling between the Br$\gamma$ and the compact centimeter emission indicates an evolving compact \hii region. This configuration points to a scenario where the central ionizing source continues to interact with the surrounding dense molecular cloud. The compact ionized core with extended molecular emission is a signature of \uchii regions that transition to more evolved {phases} as the ionized front gradually expands and disrupts the molecular cloud.

%Additionally, the presence of strong Br$\gamma$ emission in \iras, coupled with the more compact centimeter emission detected at 3 and 6~cm, is indicative of an evolving \uchii region in which the central ionizing source is still interacting with the dense, surrounding molecular cloud. The combination of a compact ionized core and extended molecular emission is a signature of \uchii regions that are transitioning into more evolved \hii regions as the ionized front expands outward, disrupting the molecular cloud.
%
The \ion{H}{1} recombination lines, on the other hand, trace the ionized gas closer to the central star. The extended Br$\gamma$ emission observed in the K-band supports the presence of a dense, ionized region surrounding the central star, with the strongest emission concentrated in the northeast direction. This morphology is consistent with the hypothesis that the central star has cleared a cavity in the molecular cloud to the NE direction, allowing ionizing radiation to escape towards the interstellar medium.

The detection of deeply embedded near-infrared objects toward the southern and western directions of the $K$-band nebulosity suggests a region of high extinction to the west. Supporting this scenario, sub-mm emission from ATLASGAL reveals an elongated structure indicative of a high column density in that direction. Together, these observations provide a comprehensive view of the molecular cloud associated with \iras, highlighting areas of dense gas and potential sites of triggered star formation as the ionization front advances into the quiescent molecular material.

\subsection{Near-infrared characterization of the nebular emission in \texorpdfstring{\iras}{IRAS 16475-4609}}

The near-infrared TripleSpec/SOAR spectroscopic observations of the near-IR counterpart of \iras provided key insights into the characterization of the nebular emission and the nature of the central source powering the \hii region in \iras.
By using by-products from the original \texttt{Spextool} IDL data reduction pipeline, we constructed spectroimages that enabled us to analyze the nebular emission associated with \iras along the slit of the TripleSpec/SOAR observations.

The analysis of the nebular emission revealed a typical \hii region spectrum, characterized by narrow emission lines of \ion{H}{1} and \ion{He}{1}, alongside significant forbidden lines such as [\ion{S}{3}] and multiple [\ion{Fe}{2}] features.
The roto-vibrational Boltzmann diagram of the \hh emission was used to constrain the properties of the excited molecular gas, leading to a rotational temperature of \trot~=~$(3.89 \pm 0.27) \times 10^3$~K and a column density of $\sim$(6.3$\pm$3.0)~$\times$~10$^{17}$~cm$^{-2}$. 
By taking advantage of the spatial information of the spectroscopic data, we identified three \hh knots (A, B, and C, Fig.~\ref{fig_h2diagram_p2p}) with distinct physical characteristics.
Knot A is located closer to the central source powering the \hii region, likely tracing regions with higher excitation conditions due to its proximity to the ionizing star. In contrast, knot B is likely a low-density region, as indicated by its lower flux relative to knots A and C. This could occur due to a combination of factors, such as the local depletion of molecular gas caused by a stronger influence of stellar winds or radiation pressure from the central source, leading to a reduced emission in this region. The lower density of \hh molecules may result in less efficient molecular hydrogen excitation, explaining the diminished \hh emission towards knot B.

\subsection{The massive star source powering the \texorpdfstring{\hii}{HII} region in \texorpdfstring{\iras}{IRAS 16475-4609}}

Based on the OB classification scheme from \citet{Hanson96,Hanson02}, the presence of broad \ion{H}{1} photospheric absorption features, along with \ion{He}{1} absorption lines, suggests that the near-IR source is an early B-type massive star. Further comparisons with spectra from known OB standards observed with TripleSpec/SOAR refined the classification to a spectral type between B0 and B0.7~V (Sect.~\ref{sec_sptype}).

We provide three independent distance estimates for \iras using spectrophotometric, astrometric, and kinematic methods (Sect.~\ref{sec_distance}).
Using available 2MASS photometry (Sect.~\ref{sec_spectrophotometric_distance}), we derived a spectrophotometric distance of 3.6$\pm$1.6~kpc to the central source powering the \hii region.
For a single-object measurement, the uncertainty associated with the spectrophotometric distance is relatively larger ($\sim$44\%) due to the error of the absolute visual magnitude calibration ($\pm$0.67~mag) from \citet{Vacca96}.
The scenario is even worse when adopting other works such as \citet{Wegner06}, where uncertainties in the absolute magnitude exceed $\pm$2~mag.
Future improvements in the fundamental parameters of OB stars can be achieved by combining data from recent all-sky surveys (e.g., Gaia~DR3) with state-of-the-art photospheric models of massive stars (e.g., Isosceles; \citealt{Araya23}).

Astrometric and photometric measurements from GaiaDR3 (Sect.~\ref{sec_gaia_distance}) yield a photogeometric distance of 3.50$^{+1.33}_{-0.89}$~kpc, which is consistent with the spectrophotometric distance and provides an independent confirmation of the distance to \iras.
Additionally, by leveraging the hydrogen recombination line measurements from SHRDS (Sect.~\ref{sec_rrl_distance}), we determined a kinematic distance of 3.47$\pm$1.26~kpc. This value is in agreement with both the spectrophotometric and Gaia-based estimates, further reinforcing the distance determination. All three distance estimates are listed in Table~\ref{table_distances}.

\begin{table}[!ht]
    \centering
    \caption{Summary of the distance estimates to \iras.}
    \label{table_distances}
    \begin{tabular}{c|l}
    \hline
    \hline
        Distance (kpc) & Method       \\
    \hline
        3.6$\pm$1.6    & Spectrophotometric (Sect.~\ref{sec_spectrophotometric_distance}) \\
        3.50$^{+1.33}_{-0.89}$ & Parallax (Sect.~\ref{sec_gaia_distance}) \\
        3.47$\pm$1.26  & Kinematic (Sect.~\ref{sec_rrl_distance}) \\
    \hline
        3.51$\pm$0.74  &  Weighted mean distance          \\
    \hline
    \end{tabular}
\end{table}

Combining these three distance estimates yields a weighted mean distance of 3.51$\pm$0.74~kpc, corresponding to a distance modulus of $\mu$~=~12.72$^{+0.41}_{-0.51}$~mag.
Considering that this is a single-object measurement, the resulting uncertainty of approximately 21\% on the weighted mean distance is relatively small for such large distances.
According to the Galactic structure models of \citet{Reid19}, such distance is broadly consistent with the expected locations of either the Scutum-Crux near arm (3.5~kpc) or the near side of the Norma arm (5.5~kpc). While spiral arm assignments remain uncertain due to variations in distance estimates and arm definitions, a potential association with the Scutum-Crux arm is suggested by the proximity of \iras to the massive stellar cluster Westerlund~1 ($\ell$=339\fdg55, $b$=$-$00\fdg40). With a well-constrained distance of 4.05$\pm$0.20~kpc \citep{Navarete22}, Westerlund~1 lies close enough to suggest that both objects may reside at similar distances, potentially within the same large-scale Galactic structure.

%%%%%%%%%%%%%%%%%%%%%%%%%%%%%%%

We estimated the mean Lyman continuum photon rate of $N_{\text{Ly}}$~=~47.36$\pm$0.06~dex based on hydrogen radio recombination line analysis (Sect.~\ref{sec_ionizing_source}). The Lyman flux falls below the ionizing flux predicted for an O9-O9.5~V star (47.56-47.90\,dex, \citealt{Martins05}), which is consistent with the classification of the near-infrared source as an early B-type star.
Furthermore, when applying the calibration of \citet{Vacca96}, the upper limit of the derived flux remains consistent within $\sim$2$\sigma$ with a B0-B0.5~V star ($\lesssim$~48.0$\pm$0.3\,dex), reinforcing our interpretation.

{In addition, the measured extent of the Brackett gamma (\brg) emission in the \hii region exhibits a radius of $\sim$16{\arcsec} (Fig.~\ref{fig_images}), corresponding to a projected length of $\sim$0.27$\pm$0.06~pc at the distance of \iras.}
According to the classification {scheme} outlined by \citet{Kurtz05}, the spatial extent of the ionized gas places \iras in a transitional phase between {ultra-compact and compact} \hii regions. This classification supports the interpretation that the source is a relatively young and still-evolving \hii region, with the ionized gas closely confined to the vicinity of the central star. Notably, the location of \iras at the boundary of a dense sub-millimeter structure places it at a critical interface between stellar feedback and the surrounding ISM, potentially influencing the next generation of star-forming activity.

\acknowledgments

{We thank the anonymous referee for their constructive comments and suggestions, which helped improve the quality and clarity of this manuscript.
The work of FN and SDP is supported by NOIRLab, which is managed by the Association of Universities for Research in Astronomy (AURA) under a cooperative agreement with the National Science Foundation. AD thanks FAPESP and CNPq for continuing support.
Based in part on observations obtained at the Southern Astrophysical Research (SOAR) telescope, which is a joint project of the Minist\'{e}rio da Ci\^{e}ncia, Tecnologia e Inova\c{c}\~{o}es (MCTI/LNA) do Brasil, the US National Science Foundation’s NOIRLab, the University of North Carolina at Chapel Hill (UNC), and Michigan State University (MSU).
Based in part on observations made at NSF Cerro Tololo Inter-American Observatory, NSF NOIRLab, which is managed by the Association of Universities for Research in Astronomy (AURA) under a cooperative agreement with the U.S. National Science Foundation.
This work has made use of data from the European Space Agency (ESA) mission {\it Gaia} (\url{https://www.cosmos.esa.int/gaia}), processed by the {\it Gaia} Data Processing and Analysis Consortium (DPAC, \url{https://www.cosmos.esa.int/web/gaia/dpac/consortium}). Funding for the DPAC has been provided by national institutions, in particular the institutions participating in the {\it Gaia} Multilateral Agreement.}

\facilities{SOAR (TripleSpec), CTIO (NEWFIRM)}

\clearpage

\bibliographystyle{aasjournal}

\appendix

\section{Comparison between TripleSpec/SOAR observations of \texorpdfstring{\iras}{IRAS 16475-4609} and OB standards}
\label{appendix_OB}

\begin{figure*}[!ht]
    \centering
    {\includegraphics[width=\linewidth]{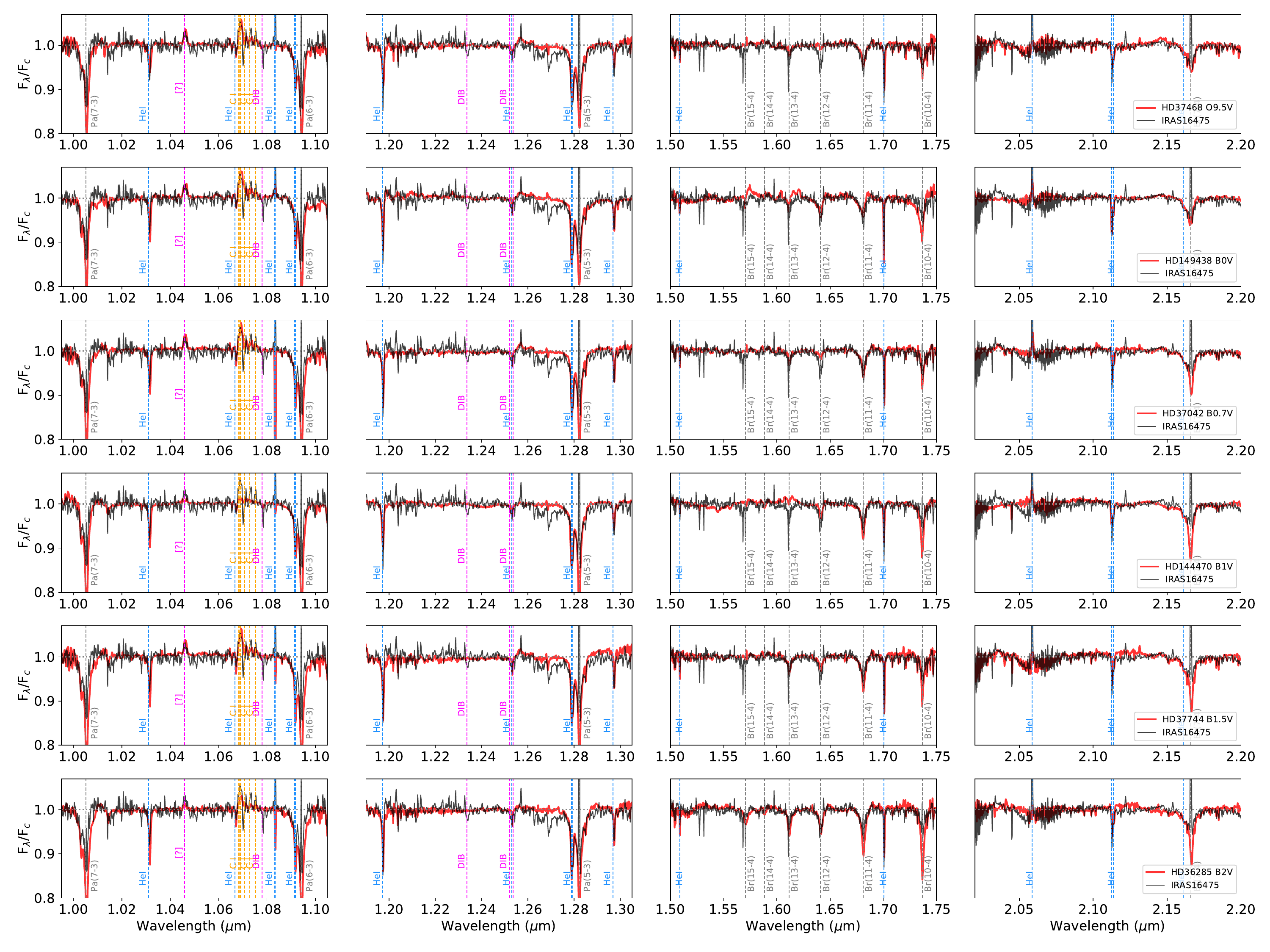}} \\[-1.0ex]
    \caption{Continuum-normalized TripleSpec spectrum of IRAS~16475-4609 used for spectral type determination.
    The red curves are a sample of representative OB-type stars observed with TripleSpec (Navarete \textit{et al.}, {in prep.}), from top to bottom: HD37468(O9.5~V), HD149438 (B0~V), HD37042 (B0.7~V), HD144470 (B1~V), HD37744 (B1.5~V), and HD36285 (B2~V). The vertical dashed lines indicate \ion{H}{1} (grey) and \ion{He}{1} (blue) lines used for classification, together with DIB and unknown features (magenta).}
    \label{fig_spec_comparisonOB}
\end{figure*}

Figure~\ref{fig_spec_comparisonOB} compares the normalized spectrum of \iras with six OB standards with spectral types ranging from O9.5~V to B2~V (from top to bottom).
Each panel exhibits useful spectral regions within the near-infrared atmospheric windows covered by the TripleSpec observations.

The first-row panels present the 1.0-1.1~{\micron} region, delimited by the Pa$\gamma$ and Pa$\beta$ lines. Three \ion{He}{1} transitions are identified in this region, together with a DIB absorption at 1.078~{\micron} and four unknown emission features in the 1.04-1.08~{\micron} range. In general, the \ion{H}{1} and \ion{He}{1} lines gets stronger towards later spectral types, while the strength of the emission features are only fainter in the spectra of the B1 and B2~V subtypes. The strength of the Pa$\beta$ at 1.09~{\micron} well-matches the B0~V spectrum.

The second-row panels show the 1.2-1.3~{\micron} window from the J band, containing four \ion{He}{1} transitions and the Pa$\beta$ feature, in addition to a few DIB absorptions. The Pa$\beta$ and \ion{He}{1}$\lambda$12789/94{\AA} doublet matches the B0~V and B0.7~V spectra.

The third-row panels exhibit the 1.5-1.75~{\micron} region from the H band. Several high-order transitions of the \ion{H}{1} Brackett series are located within this region, together with two \ion{He}{1} transitions. The higher-order Brackett series are problematic in the comparison spectra, however the first orders are generaly reasonably well-sampled. The wings of the \ion{H}{1} lines from the \iras spectrum are well matched by the B0.7~V star, but also for the earlier O9.5~V.

The forth-row panels exhibit the 2.02-2.20~{\micron} region of the K-band window, containing the \brg feature and two \ion{He}{1} lines. Similarly as observed in the H-band, the wings of the \brg is consistent with spectral types ranging from O9.5-B0.7~V.

\end{document}